%% file: UVMEM_GPU_Paper.tex
\documentclass[twocolumn, 5p, authoryear]{elsarticle}
\usepackage{graphicx}	% Including figure files
\usepackage{amsmath}	% Advanced maths commands
\newcommand{\vect}[1]{\boldsymbol{#1}}
\usepackage{booktabs}
\usepackage{amssymb}	% Extra maths symbols
\usepackage{multicol}        % Multi-column entries in tables
\usepackage{bm}		% Bold maths symbols, including upright Greek
\usepackage{pdflscape}	% Landscape pages
\usepackage{algorithm}
\usepackage{algorithmic}
\usepackage{url}
\usepackage{hyperref}
\usepackage{subfig}
\usepackage{setspace}
\usepackage{color}
\usepackage{multirow}
\usepackage[english]{babel}
\usepackage{hyphenat}
\usepackage[normalem]{ulem}

\begin{document}
\selectlanguage{english}

\title{Multi-GPU maximum entropy image synthesis for radio astronomy}

\author[diinf]{M.~C\'arcamo}
\ead{miguel.carcamo@usach.cl}
\author[diinf,cmm]{P.E.~Rom\'an}
\author[das]{S.~Casassus}
\author[das]{V.~Moral}
\author[diinf]{F.R.~Rannou}

% List of institutions

\address[diinf]{Departamento de Ingenier\'ia Inform\'atica, Universidad de Santiago de Chile, Av. Ecuador 3659, Santiago, Chile}
\address[cmm]{Center for Mathematical Modeling, Universidad de Chile, Av. Blanco Encalada 2120 Piso 7, Santiago, Chile}
\address[das]{Astronomy Department, Universidad de Chile, Camino El Observatorio 1515, Las Condes, Santiago, Chile}

\input{abstract}

\maketitle

\input{introduction}

\input{methods-implementation}

\input{materials-methods}
\input{results}
\input{conclusion}
\section*{Acknowledgements}
This paper makes use of the following ALMA data: ADS/JAO.ALMA \#2011.0.000015.SV, ADS/JAO.ALMA \#2011.0.00465.S, ADS/JAO.ALMA \#2011.0.00003.SV, ADS/JAO.ALMA \#2011.0.000016.SV, ADS/JAO.ALMA \#2013.1.00305.S. 
ALMA is a partnership of ESO (representing its member states), NSF (USA) and NINS (Japan), together with NRC (Canada), NSC and ASIAA (Taiwan), and KASI (Republic of Korea), in cooperation with the Republic of Chile. The Joint ALMA Observatory is operated by ESO, AUI/NRAO and NAOJ. Also, the calculations used in this work were performed in the Brelka cluster, financed by Fondequip project EQM140101 and housed at MAD/DAS/FCFM, Universidad de Chile. F. R. Rannou and M. Carcamo were partially funded by DICYT project 061519RF, Universidad de Santiago de Chile. P. Roman acknowledges postdoctoral FONDECYT projects 3140634, Basal PFB-03 Universidad de Chile, FONDECYT grant 1171841, and Conicyt PAI79160119 Universidad de Santiago de Chile. S. Casassus acknowledges support from  Millennium Nucleus RC130007 (Chilean Ministry of Economy), and additionally by FONDECYT grant 1171624.

% Entry for the table of contents, for this guide only
\addcontentsline{toc}{section}{Acknowledgements}
%%%%%%%%%%%%%%%%%%%%%%%%%%%%%%%%%%%%%%%%%%%%%%%%%%
%%%%%%%%%%%%%%%%% APPENDICES %%%%%%%%%%%%%%%%%%%%%
\appendix
\input{appendix}


%\clearpage
\bibliographystyle{model2-names}
\bibliography{bibliography} % if your bibtex file is called example.bib

\end{document}

%% file: abstract.tex
% !TEX root = ./UVMEM_GPU_Paper.tex
% !BIB program = bibtex

\begin{abstract}
 The maximum entropy method (MEM) is a well known deconvolution technique in radio-interferometry. This method solves a non-linear optimization problem with an entropy regularization term. Other heuristics such as CLEAN are faster but highly user dependent. Nevertheless, MEM has the following advantages: it is unsupervised, it has a statistical basis, it has a better resolution and better image quality under certain conditions. This work presents a high performance GPU version of non-gridding MEM, which is tested using real and simulated data. We propose a single-GPU and a multi-GPU implementation for single and multi-spectral data, respectively. We also make use of the Peer-to-Peer and Unified Virtual Addressing features of newer GPUs which allows to exploit transparently and efficiently multiple GPUs. Several ALMA data sets are used to demonstrate the effectiveness in imaging and to evaluate GPU performance. The results show that a speedup from 1000 to 5000 times faster than a sequential version can be achieved, depending on data and image size. This allows to reconstruct the HD142527 CO(6-5) short baseline data set in 2.1 minutes, instead of 2.5 days that takes a sequential version on CPU.
\end{abstract}

%% file: introduction.tex
% !TEX root = ./UVMEM_GPU_Paper.tex
% !BIB program = bibtex

\section{Introduction}

Current operating radio astronomy observatories (e.g. ALMA, VLA, ATCA) consist of a number of antennas capable of collecting radio signals from specific sources. Each antenna's signal is correlated with every other signal to produce samples of the sky image $I(x,y)$, but on the Fourier domain \citep{fft}. These samples $V(u,v)$ are called visibilities and comprise a sparse and irregularly sampled set of complex numbers in the $(u,v)$ plane. A typical ALMA sampling data set contains from $10^4$ to more than $10^9$ sparse samples in one or more frequency channels.

In the case where $V(u,v)$ is completely sampled, Equation \ref{eq:fft} states the simple linear relationship between image and data: 
\begin{align}
\label{eq:fft}
V(u,v) = \int_{\mathbb{R}^{2}} A(x,y)I(x,y)e^{-2\pi i(ux+vy)}dxdy
\end{align}

Thus the image can be recovered by Fourier inversion of  the interferometric signal \citep{clark}. In this equation, kernel $A(x,y)$ is called the primary beam (PB) and corresponds to the solid angle reception pattern of the individual antennas and it is modelled as a Gaussian function. If the antennas are dissimilar, it is the geometric mean of the patterns of the individual antennas making up each individual baseline \citep{libroAstro2}. 

\begin{figure}
\centering\includegraphics[width=0.8\linewidth]{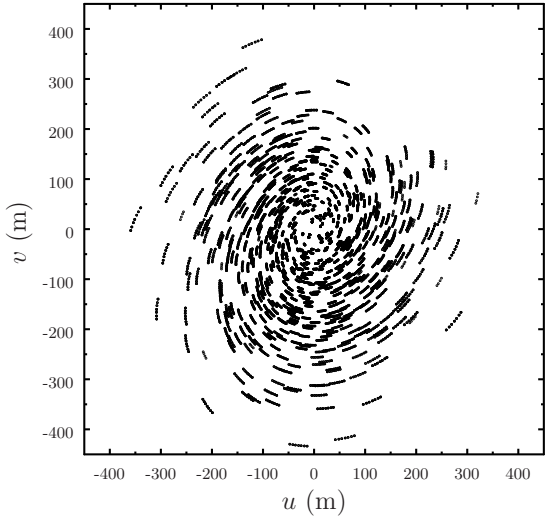}
\caption{Short baseline $uv$ sampling of the HD142527 protoplanetary disk.}
\label{fig:planouv}
\end{figure}

In the real scenario of collecting noisy and irregularly sampled data, this problem is not well defined \citep{marechal,illposed}. To approximate the inverse problem of recovering the image from a sparse and irregularly sampled Fourier data a process called Image Synthesis \citep{libroAstro} or Fourier Synthesis \citep{marechal} is used. Current interferometers are able to collect a large number of (observed) samples in order to fill as much as possible the Fourier domain. As an example, Figure \ref{fig:planouv} shows the ALMA 400 meter short baseline sampling for Cycle 2 observation of the HD142527 plotoplanetary disc. Additionally,  the interfe\-rometer is able to estimate data variance $\sigma^{2}_k$ per visibility as a function of the antenna thermal noise \citep{libroAstro}.

%%
%%% FIGURE 1: FOURIER SAMPLING
%%

Many algorithms have been proposed for solving the image synthesis problem and the standard procedure is the CLEAN heuristic \citep{hogbom}. This algorithm is based on the dirty image/beam representation of the pro\-blem (\ref{apx:beam}), which results in a deconvolution problem \citep{libroAstro2}. CLEAN has been interpreted as a matching pursuit heuristic \citep{persuit}, which is a pure greedy-type algorithm  \citep{greedyAlgorithm}. Image reconstruction in CLEAN is performed in the image space using the convolution relationship, and it is therefore quite e\-ffi\-cient\-ly implemented using FFTs. This algorithm is also supervised. The user could indicate iteratively in which region of the image the algorithm should focus. However, statistical interpretation of resulting images and remaining artifacts are yet to be described by a well founded theory.

MEM is inspired by a maximum likelihood argumentation since interferometer measurements are assumed to be corrupted by Gaussian noise \citep{optimal_image,VoronoiCabrera}. Reconstructed images with this method have been considered to have higher re\-so\-lu\-tion and fewer artifacts than CLEAN images \citep{smemda,MEM,nearlyBlack}. MEM  was the second choice image synthesis algorithm \citep{firstVersionMEM}, and is mainly used for checking CLEAN bias and images with complex structure \citep{astromem1,astromem2,astromem3}. However, routine use of MEM with large data sets has been hindered due to its high computational demands \citep{mem1, libroAstro2, MEM}. For instance, to reconstruct an image of  $10^{3} \times 10^{3}$ pixels from the ALMA Long Baseline campaign dataset HL Tau Band 3 (279,921,600 visibilities), MEM computes $7.9 \times 10^{16}$ floating-point operations per iteration, approximately. It was also claimed that MEM could not reconstruct point-source like images over a plateau e.g. \citep{memGeneralized}. However, we have found (Section \ref{subsec:imagingres}) this not to be true using a simulated data set. In fact, the experiments  show MEM can reconstruct the point source and maintain a smooth plateau.

The MEM algorithm is traditionally implemented with optimization methods (OM) based on the gradient \citep{mem1}, also called first order methods. 

Typical examples of such methods are  the conjugate gradient \citep{numericalrecipes}, quasi-newton methods like the modified Broyden's L-BFGS-B \citep{numericaloptimization}, and the Nesterov's class of accelerated gradient descent \citep{ConvexOptimization}. All of them require the first derivative calculation which has a computation complexity of $O(M\cdot Z)$, where $Z$ is the number of samples and $M$ is the number of image pixels. The gradient computation is the most expensive part per iteration of first order OMs. In consequence, such OMs are equivalent per iteration from the complexity point of view. In particular, we choose to study the GPU implementation for the Conjugate Gradient (CG).  This version uses a positive projected gradient due its simplicity for the case of image synthesis and large data ($Z\sim 10^{9}$ and $M\sim 10^{7}$).

A GPU implementation that supports Bayesian Inference for Radio Observations (BIRO) \citep{BIRO} has been proposed before \citep{montblanc}. This approach uses Bayesian inference to sample a parameter set re\-pre\-sen\-ting a sky model to propose visibilities that best match the model.
Their model handles point and Gaussian sources. Our approach employs a non-linear optimization problem to directly solve the Bayesian model with a maximum entropy prior using real and synthetic data. However, our method is not currently able to estimate uncertainties by itself as BIRO does.

In this paper we present a high performance, high-throughput implementation of MEM for large scale, multi-frequency image synthesis. Our aim is to demonstrate computational performance of the algorithm making it practical for research in radio-interferometry for today data sets. The main features of the solution are:

\begin{itemize}
    \item \textbf{GPU implementation}: With the advent of larger interferometric facilities such as SKA \citep{ska} and LOFAR \citep{lofar}, efficient image synthesis based on optimization cannot be delivered by modern multi-core computers. However, we have found that the image synthesis pro\-blem, as formulated by MEM, fits well into the Single Instruction Multiple Thread (SIMT) paradigm (Section \ref{subsec:gpu}), such that the solution can make efficient use of the massive array of cores of current GPUs. Also, our GPU proposal exploits features like the Unified Virtual Addressing (UVA) and the \textit{Peer-to-peer} (P2P) communications to devise a multiple GPU solution for large multi-frequency data.
    \item \textbf{Parameterized reconstruction}: Supervised algorithms could become obsolete in this high-throughput regime in favor of more automatic methods based on a fitting criteria. Although MEM requires a few parameters, it does not require user assistance during algorithm iteration. In this sense, we say MEM is an unsupervised image synthesis algorithm.
    \item \textbf{Non-gridding approach}: Gridding is the process of resampling  visibility data into a regular grid \citep{libroAstro2}. This task is usually carried out by convolving the data with a suitable kernel. However, the resampling process could produce biased results: loose of flux density \citep{gridding}, aliasing and Gibb's phenomenon \citep{ImagingBriggs}. Although this processing reduces the amount of data and the computation time by data averaging, it does to the detriment of further statistical fit. The use of a high performance implementation allows us to process the full dataset without gridding, and still achieve excellent computational performance. However, an interpolation step is still needed to compute model visibilities from the Fourier transform of the image estimate.
    \item \textbf{Mosaic support}: Mosaic images in interferometry allow the study of large scale objects in the sky. In this Bayesian approach we fit a single model image to the ensemble of all pointings. The process of image restoration (see Sec.\ref{sec:restoration}) is then performed on residual images obtained with the linear mosaic formula \citep{libroAstro,libroAstro2}. 
    \item \textbf{Multi-frequency support}: Spectral dependency can introduce strong effects into image synthesis \citep{MULTIFREQUENCY}. In this implementation we have applied our algorithm to reconstruct ALMA datasets with several channels and spectral windows, but for a single band. Typically, total bandwidth amounts to a maximum of $\sim$10\% the central frequency. In these cases, image synthesis can be performed assuming a zero spectral index \citep{libroAstro}. 
\end{itemize}

%% file: methods-implementation.tex
% !TEX root = ./UVMEM_GPU_Paper.tex
% !BIB program = bibtex

\section{Method and implementation}

This section describes the mathematical formulation of the mono-frequency \textit{Maximum Entropy Method} (MEM) and the multi-frequency MEM together with a positively constrained conjugate gradient minimization algorithm. Finally, the GPU implementation details are given.

\subsection{Data description}

Let $V^o=\{V^o_k\}$, $k=0,\ldots, Z-1$, be the (observed) visibility data, let $\{\sigma_k\}$ $k=0,\ldots, Z-1$ the estimated deviations of each sample, and let $I=[I_i]$, $i=0,\ldots,M-1$.
be the image to be reconstructed from $V^o$. Notice that $I$ is a regularly sampled, and usually 
a square image function, whereas $V^o$ is a list of sparse points.

Sampled visibilities are grouped in spectral windows. Spectral windows are contiguous spectrum whose frequencies are uniformly spaced and also uniformly divided in channels. Therefore, for the case of multi-spectral, multi-frequency data the observed visibilities are indexed as $V^o_{w,c,k}$ where $w$ corresponds to a spectral window, and $c$ corresponds to a channel of the spectral window $w$. 

Interferomeric data also contains polarization. We simplify our approach by considering only the case of intensity polarization (I). This simplification is rather general in practice. Radio interferometers such as ALMA, VLA, or ATCA generate data on orthogonal polarizations (either linear XX, YY, or circular LL, RR), which can enter directly in the algorithm as polarization I e.g. \citep{libroAstro2}. 

\subsection{Bayes formulation and the maximum entropy method}
\label{subsec:mem1}

The Maximum Entropy Method (MEM) can be seen as a Bayesian strategy that selects one image among many feasible. Since the data is noisy and incomplete, the solution space is first reduced choosing those images that fit the measured visibilites to within noise level.  Among these, the Bayesian strategy selects the one that has a maximum probability of being observed according to a counting rule \citep{cornwell, MEM, libroAstro}.

Specifically, let $P(V^o|I)$ be the likelihood of observing the data  given image $I$ and let $P(I)$ be  a \emph{prior} knowledge of the image based on the chosen counting rule. Then, by Bayes theorem we obtain the \emph{a posteriori} pro\-ba\-bi\-li\-ty \citep{1993Pixon}

\begin{align}
\label{eq:BayesArg}
P(I|V^o) = \frac{P(V^o|I)P(I)}{P(V^o)}
\end{align}
where $P(V^o)$ is the normalizing constant. Thus, the Maximum a Posteriori (MAP) image estimate is

\begin{align}
\hat{I}_{\text{MAP}}  = \arg \max_{I} P(I|V^o)
\label{eq:map1}
\end{align}

$\hat{I}_{\text{MAP}}$ corresponds to that image with  maximum probability of occurring according to the assumed prior knowledge, among those with maximum probability of producing the observed data within noise level. 
As exponential functions will be used to model the likelihood, it is preferred to work with the logarithm of Equation (\ref{eq:map1}). Dropping term $\log P(V^o)$, independent of $I$, we obtain
\begin{align}
\hat{I}_{\text{MAP}}  = \arg \max_{I} \Phi(I, V^o) 
\label{eq:map2}
\end{align}
where
\begin{align}
\Phi(I, V^o) = \log P(V^o|I) + \log P(I)
\label{eq:phi}
\end{align}

The likelihood $P(V^o|I)$ can be approximated using the fact that visibilities are independent and identically distributed Gaussian random variables corrupted by Gaussian noise of mean zero and standard
deviation $\sigma_k$. Then, the likelihood can be expressed as

\begin{align}
\label{eq:DataProbability}
    P(V^o|I) \propto \displaystyle\prod_k \exp\biggl\{-{\frac{1}{2}\biggl|\frac{V^m_k(I) - V^o_k}{\sigma_k}}\biggr|^2\biggr\}
\end{align}
where $V^m(I)$ denotes model visibilities which are functions of the image estimate. 

The MEM image prior is assumed to be a multinomial distribution \citep{NatureMEM} of a discrete total intensity that covers the entire image, in analogy with a CCD camera array receiving quantized luminance from the sky \citep{1993Pixon}. Let $M$ be the number of image pixels and let $N_i=I_i/G$ be the quantized brightness \citep{optimal_image} collected at pixel $i$. $G$ is considered as the minimal indistinguishable signal variation \citep{1993Pixon}. Then, the prior can be expressed by counting equivalent brightness configurations:

\begin{align}
\label{eq:entropy}
P(I) = \frac{N!}{M^N\prod_i{N_i!}} 
\end{align}
where $N = \sum_i N_i$. Under the assumption of a large number of quantas per pixel the Stirling's approximation ($\log(x!)\sim x\log(x) -x, x\gg 1$), which results in equation (\ref{eq:logentropy}). There is a subtle difference from derivation found in \citep{optimal_image,1993Pixon} since the quantization scale $G$ is explicit in the entropy \citep{VoronoiCabrera}. This results in an additional constraint on the sky intensity $I_i\geq G$, where $G$ is sufficiently small to reproduce a large number of brightness quantas in signal. Taking logarithms to Equations (\ref{eq:DataProbability}) and (\ref{eq:entropy}), we obtain

\begin{eqnarray}
\label{eq:loglikelihood}
\log(P(Ṿ^o|I)) = & - \frac{1}{2}\sum_{k}{{\biggl|\frac{ V^m_{k}(I) -  V^o_k}{\sigma_k}\biggr|^2}} \\
\log(P(I)) \sim S  = & -\sum_i{\frac{I_i}{G} \log{\frac{I_i}{G}}} + \text{constant} \label{eq:logentropy}
\end{eqnarray}

Replacing Equations (\ref{eq:loglikelihood}) and (\ref{eq:logentropy}) into Equation (\ref{eq:phi}), we arrive at the following objective function for minimization with $\lambda=1$:

\begin{align}
\label{eq:phi2}
 \Phi(I,V^o;\lambda, G) = \frac{1}{2}\sum_{k}{{\biggl|\frac{V^m_{k}(I) - V^o_k}{\sigma_k}\biggr|^2} + \lambda \sum_i{\frac{I_i}{G} \log{\frac{I_i}{G}}}}
\end{align}

We recognize Equation (\ref{eq:phi2}) as a typical $\chi^2$ term plus an entropy regularization term $-\lambda S$ with a penalization factor $\lambda$ as in (\ref{eq:phi3}). The previous generalization is a common expression of the objective function for MEM \citep{MEM}. Other authors argument penalization in equation (\ref{eq:phi2}) as an entropy-based distance from $I$ to a prior blank image $G$ \citep{mem1}. 

\begin{equation}
\label{eq:phi3}
\Phi(I, V^o; \lambda, G) =  \chi^2(V^{o},I) - \lambda S(I; G)
\end{equation}

For multi-frequency data, the $\chi^2$ term is expressed as: 

\begin{align}
\label{eq:chi2-multi}
\chi^2 = \frac{1}{2}\sum_{w=0}^{W-1}\sum_{c=0}^{C-1}\sum_{k=0}^{Z-1}{{\biggl|\frac{V^m_{w,c,k}(I) - V^o_{w,c,k}}{\sigma_{w,c,k}}\biggr|^2}}
\end{align}

Finally, the MEM method for image synthesis consists of solving the following non-linear, constrained  optimization problem:
\begin{equation}
\hat{I}_{\text{MEM}} =  \arg \min_{I\geq G} \Phi(I;V^o; \lambda, G)
\label{eq:minproblem}
\end{equation}

\subsection{MEM parameters}

Although our implementation of MEM is an unsupervised algorithm it is still dependent on four important parameters that determine properties of the resulting image.
These parameters are the entropy penalization factor $\lambda$, the minimal image value $G$, pixel size $\Delta x$, and image size $M^{1/2}$. Even though a complete study on how these parameters affect image properties is beyond the scope of this work, we briefly mention our approach to select reasonable values for them.

In theory, pixel size should satisfy Nyquist criterion, such that image sampling rate should be at least twice as large as the maximum sampled frequency, that is $1/\Delta x \ge 2 u_\text{max}$. In practice, this is approximated as 1/5  to 1/10 of the Full-Width-Half-Maximum (FWHM) across the main lobe of the dirty beam (see \ref{apx:thermal}).  

Assuming the primary beam entirely covers the region of interest, it is re\-co\-mmen\-ded that the canvas size ($M^{1/2} \Delta x$) has to be larger than the full extent of the main signal. This argumentation is based on the fact that MEM is known to have optimal performance for nearly black images \citep{nearlyBlack}.

In the absence of a non-blank prior image, it is assumed that the minimal value a pixel can take ($G$) is at most the thermal noise $\sigma_D$ given by (see \ref{apx:thermal})
\[ \sigma_D = \frac{1}{\sqrt{\sum_k \frac{1}{\sigma^2_k}}} \]
However, it is preferred to use a small fraction of $\sigma_D$, specially in low signal-to-noise regimes. Notice that when (and if) the restored image is computed, an additional factor of $\sigma_D$ is added to the image.

The penalization factor $\lambda \geq 0$ controls the relative importance between data and the entropy term. When $\lambda=0$ the problem becomes a least-square optimization problem. In case of a very small lambda the problem is nearly a least-square with a lower bound constrain ($ I \geq G$). But when $\lambda$ increases, less importance is given to data and solution becomes smoother until become a constant equal to $G$.
As long as the image synthesis problem has a non-unique solution, each value for $\lambda$ represents a possible solution for the problem. Therefore, MEM allows to explore smoother solutions by changing this single parameter at the cost of degrading residuals ($\chi^2$). For example, we start by choosing a very small value of $\lambda \approx 10^{-6}$ which can be considered a near least-square solution. Then, we re-run the program increasing $\lambda$ by a factor of ten, until the user is satisfied. This procedure can be made practical with a fast enough algorithm implementation.

\subsection{Objective function evaluation}

The minimization problem (\ref{eq:minproblem}) can be solved by a cons\-trained Conjugate Gradient (CG) algorithm, which repeatedly evaluates the objective function and its gradient to compute the search direction and step size. Evaluation of the entropy term is straightforward, but evaluation of the $\chi^2$ term requires some additional processing. The following steps are required for computation of model visibilities $V^m$ required for evaluating $\chi^2$.
\begin{enumerate}
    \item The attenuation image $A_{m,n}$ or primary beam represents the discrete version of the reception pattern of the telescope in Equation \ref{eq:fft}. This image is generally modeled as a Gaussian and depends on the radio-interferometer \citep{libroAstro2} scaling linearly with the frequency. For ALMA the FWHM is 21 arcsec at 300GHz for antennas of 12m diameter. 
    \item Model visibilities $V^F$ on a grid are obtained by applying a 2D Discrete Fourier Transform to the model image. For the case of multi-frequency data, an atte\-nua\-tion matrix $A_{w,c}$ is built for each channel of each spectral window, resulting in a set of model visibilities, $V^F_{w,c}$. Thus, the Fourier modulation and the interpolation step are carried out independently for each frequency channel.
    \begin{equation*}
    V^F_{w,c} = \mathcal{F}_{\text{2D}}\{ A_{w,c} \cdot I\}
    \end{equation*}
    In case of a mosaic measurement, for each pointing in the sky an attenuation image is calculated and centered in the corresponding field of view. 
    \item The phase-tracking of the object has a center according to the celestial sphere \citep{libroAstro2} coordinates. This implies that the image is shifted according to that center. Thus, a modulation factor is applied as follows:

    \begin{equation*}
        \hat{V}^F(u,v) = V^F(u,v) \exp\{2\pi i(ux_c + vy_c)/M^{1/2}  \} 
    \end{equation*}

    where $(u,v)$ are the uniformly spaced grid locations, and $(x_c, y_c)$ are the direction cosines of the phase-tracking center. In case of mosaic image, data is divided into several pointing in the sky shifting each field of view a\-ccor\-ding to its location.
    \item Model visibilities $V^m$ are approximated from $\hat{V}^F$ by using bilinear interpolation method. Bi\-li\-near interpolation has the effect of local smoothing over a cell grid, thus avoiding overshoot effects at object's edges \citep{imageProcessing}. Aliasing is still an issue in this method, but a common way to overcome this problem is by reducing the Fourier grid cell size increasing the number of image pixels. 
    \item Residual visibilities $V^R=\{V^R_k\}$ are calculated from $V^R_k =V^m_k - V^o_k$ in order to compute the $\chi^2$ term.
\end{enumerate}

It is worth emphasizing that our algorithm computes the error term for each visibility at their exact $uv$ locations, and do not apply any type of Fourier data gridding to reduce the amount of computation.

\subsection{Gradient evaluation}

Gradient computation is required for first order optimization methods. Notice that before gradient function is called, the objective function is always executed. Thus, residual visibilities $V^R$ are available for gradient calculation. As well as the objective function the first computation has to do with the entropy gradient. Therefore, every coordinate of the entropy gradient ($\nabla S$) has a value that follows Equation \ref{eq:Sgradient}.

\begin{equation}
    [\nabla S]_i = 1 + \log{\frac{I_i}{G}}
\label{eq:Sgradient}
\end{equation}

The $\chi^2$ gradient term is calculated from the derivative of the Direct Fourier Transform of the image (see \ref{apx:chi2gradient}) as shown in Equation (\ref{eq:chi2gradient}). The term $X_i = (x,y)$ is the image coordinate of the $i$th pixel according to the direction cosines of the phase tracking center, and $U_{w,c,k} = (u,v)$ is the sampling coordinate of the $k$th visibility at channel $c$ of spectral window $s$.

\begin{equation}
[\nabla \chi^2]_{m,n} = \sum_{k=0}^{Z-1} \frac{\text{Re}\bigl(V^R_k e^{2\pi i X_i\cdot U_{k}}\bigr)}{\sigma^2_k}
\label{eq:chi2gradient}
\end{equation}

\subsection{SIMT and GPU implementation} 
\label{subsec:gpu}

The Single Instruction Multiple Thread (SIMT) paradigm is a compute model in which one instruction is applied to several threads  in parallel. This means that when a group of threads is ready to run, the SIMT scheduler broadcasts the next instruction to all threads in the group.  This is similar but not equal to the Single Instruction Multiple Data (SIMD) paradigm, in which the same instruction is  executed over different data, without referring to a particular execution model. This distinction is subtle but important to produce efficient GPU solutions, because an understanding of what is involved in terms of thread execution can make a big difference in the design and performance of a code. 

In particular, our approach satisfies the following three general principles for effective and efficient GPU solutions:

\begin{enumerate}
    \item Write simple and short kernels to use a small number of registers per thread, increasing warp number per Streaming Multiprocessor.
    \item Write kernels with spatial locality access to increase memory transfer and cache hit rate.
    \item Keep most of the data in global memory to avoid slow transfers from host to device and vice versa.
\end{enumerate}

Although a complete discussion of GPU programming and GPU terminology can be found in \citep{cudapro}, we briefly
discuss here some of the terms used in this paper.  A kernel is the code executed by threads in parallel on a GPU. How and when these threads
execute on the GPU depend on how the programmer decomposes the problem into sub-problems that can be solved
in parallel on the available GPU resources. The most basic GPU resources are the processing elements or cores.
A core can run at most one thread at a time. Cores are grouped into one or more Streaming Multiprocessors (SM), and all threads in an SM execute the same instruction at the same time. From the programming point of view, threads are grouped into blocks, which in turn are organized into a grid of blocks. Different thread blocks can be scheduled to run in different SM, but once a block is assigned to a SM, all threads from that block will execute on that particular SM until they finish. 
This also means
that  all block resources, like thread registers and shared memory need to be allocated for the entire block execution. Using too many
resources per thread will limit the number of thread blocks that can be assigned to a SM, which in turn will limit the
number of active blocks running in the SM.

We have implemented a positively constrained CG algorithm to solve Equation (\ref{eq:minproblem}). The algorithm's main iteration loop is kept in host, while  compute intensive functions are implemented in GPUs. To minimize data transfer, most vectors and matrices are allocated and kept in device global memory. Only the image estimate is transferred back and forth between host and device. At convergence, model visibilities are also sent back to host to compute final residuals.

As an example, here we show implementation details of the two most compute intensive functions, namely  $\Phi$ and $\nabla\Phi$.
Algorithm \ref{alg:chi2gpu} shows the $\chi^2$ host function that invokes 1D and 2D kernels to accomplish its goal. Lines 2 to 5 apply correction factors and the Fourier transformation steps.
Although not shown here, each kernel invocation has an associated grid on which threads are run. For instance, the attenuation and modulation kernels on line (2) and (4), respectively, use a $M^{\frac{1}{2}} \times M^{\frac{1}{2}}$ kernel grid, while the interpolation kernel at line 5 uses a 1D grid of $T$ threads, where $T$ is the next higher power of 2, greater than or equal to $Z$. All 2D grids are organized in blocks of $32 \times 32$ threads, while 1D grids  are organized in blocks of $1024 \times 1$ threads.
The last two steps correspond to the residuals vector computation (line 6) and its global reduction sum (line 7).

\begin{algorithm}
	\begin{algorithmic}[1]
	    \STATE{\textbf{Chi2}$(\vect{\widetilde{I}}, \vect{V}^o, \vect{\omega}, x_c, y_c, \Delta u, \Delta v)$}
	    \STATE{$\vect{I}_a = \textbf{KAttenuation}(\vect{I}, \vect{A})$}
		\STATE{$\vect{V}^F = \textbf{KcudaFFT}(\vect{I}_{a})$}
		\STATE{$\vect{\hat{V}}^F = \textbf{KModulation}(\vect{V}^F, x_c, y_c)$}
	    \STATE{$\vect{V}^{m} = \textbf{KInterpolation}(\vect{\hat{V}}^F, u/\Delta u, v/\Delta v)$}
		\STATE{$\vect{C} = \textbf{KChi2Res}(\vect{V}^m, \vect{V}^o, \vect{\omega})$}
		\STATE{$\textbf{KReduce}(\vect{C})$}

	\end{algorithmic}
	\caption{$\chi^{2}$ host function}
	\label{alg:chi2gpu}
\end{algorithm}

\begin{algorithm}
	\begin{algorithmic}[1]
	    \STATE{\textbf{KChi2Res}($\vect{V}^m$, $\vect{V}^o$, $\vect{\omega}$)}
		\STATE{$k = blockDim.x * blockIdx.x + threadIdx.x$}
		\IF{$k < Z$}
			\STATE{$\vect{V}^{R}_k = \vect{V}^{m}_k - \vect{V}^{o}_k$}
		    \STATE{$\vect{\chi^{2}}(k) = \vect{\omega}(k) \cdot (\text{Re}(\vect{V}^{R}(k))^{2} + \text{Im}(\vect{V}^{R}(k))^{2})$}
		\ENDIF
	\end{algorithmic}
	\caption{$\chi^{2}$ 1D Kernel}
	\label{alg:chi2}
\end{algorithm}

Algorithm \ref{alg:chi2} shows the \texttt{KChi2Res} kernel. First, each thread computes  its index into the grid (line 2), and then
computes the corresponding contribution to the $\chi^2$ term (lines 4 and 5).

Algorithm \ref{alg:dchi2} lists the pseudo-code for the calculation of the gradient from Equation \ref{eq:chi2gradient}, which is the most compute intensive kernel. This kernel uses a $M^{1/2} \times M^{1/2}$ grid. The first lines (2 and 3) compute the index into the 2D grid. Then every thread calculates their corresponding contribution to a cell of the gradient, for every visibility. Recall that sine and cosine functions are executed in the special function units (SFU) of a GPU. Therefore, if Tesla Kepler GK210B has 32 SFU and 192 cores per SM, an inevitable hardware bottleneck occurs.

\begin{algorithm}
	\begin{algorithmic}[1]
	    \STATE{\textbf{KGradChi2}$(\vect{V}^R, \vect{\omega}, x_c, y_c, \Delta x, \Delta y)$}
		\STATE{$j = blockDim.x * blockIdx.x + threadIdx.x$}
		\STATE{$i = blockDim.y * blockIdx.y + threadIdx.y$}
		\STATE{$x = (j-x_{c})\cdot \Delta x$}
		\STATE{$y = (i-y_{c})\cdot \Delta y$}
		\STATE{$\vect{\nabla \chi^{2}}(N\cdot i+j) = 0$}
		\FOR{$k=0$ \TO $Z-1$}
		\STATE{$\vect{\nabla \chi^{2}}(N\cdot i+j) \mathrel{+}= \vect{\omega}(k) * [\text{Re}(\vect{V}^{R}(k)) \cdot \cos(2 \pi \langle(u_k,v_k),(x,y)\rangle ) - \text{Im}(\vect{V}^{R}(k)) \cdot \sin(2 \pi \langle(u_k,v_k),(x,y)\rangle)]$}
		\ENDFOR
	\end{algorithmic}
	\caption{$\nabla \chi^{2}$ 2D Kernel}
	\label{alg:dchi2}
\end{algorithm}

\subsection{Multi-GPU Strategy}

The main idea of using several GPUs comes from Equation \ref{eq:chi2-multi}. Since the contribution of every channel to the $\chi^2$ term  can be calculated independently from any other cha\-nnel, it is possible to schedule different channels to different devices, and then do the sum reduction in host.  Generally, the number of channels is larger than the number of devices and it is therefore necessary to distribute channels as equal as possible to achieve a balanced workload.

%Multiple GPU devices available in the same system can work together or be used independently. 
Kernels executing on 64-bits systems and on devices with compute capability 2.0 and higher (CUDA version higher than 4.0 is also required) can use \textit{Peer-to-peer} (P2P) \citep{cudapro} to reference a pointer whose memory pointed to has been allocated in any other device connected to the same PCIe root node. This, together with the Unified Virtual Addressing (UVA) \citep{cudapro}, which maps host memory and device global memory into a single virtual address space can be combined to access memory on any device transparently. In other words, the programmer does not need to explicitly program memory copies from one device to another. However, since  memory re\-fe\-ren\-cing under P2P and UVA has to be done in the same process, OpenMP is used \citep{openmp} to create as many host threads as devices to distribute channels. 
%parallelizing the $\chi^{2}$ and $\nabla \chi^{2}$ calculations.

%%
%% FIGURE 2: ILLUSTRATION OF THE MULTI-THREAD...
%%
\begin{figure}
    \centering
    \includegraphics[width=0.9\linewidth]{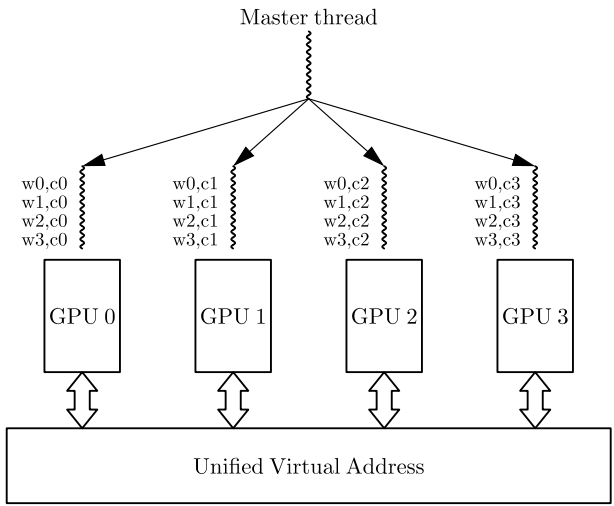}
    \caption{Illustration of the multi-thread strategy for multi-GPU reconstruction. The figure shows an example of how spectral windows (w0,w1,w2,w3) and channels (c0,c1,c2.c3) are distributed with a round-robin scheduling. Data is copied to the UVA before kernel invocation.}
    \label{fig:openmp_gpu}
\end{figure}

Figure \ref{fig:openmp_gpu} illustrates the multi-GPU reconstruction s\-tra\-te\-gy for the case of data with four spectral windows and four channels each, on a four device system. A master thread creates a team of four worker threads which are assigned channels in a round-robin fashion with portions of one channel at a time. Channel data is copied from host to the co\-rres\-pon\-ding device by the master thread. Algorithm \ref{alg:chi2openmp} shows the multi-frequency host function pseudo code that implements this idea. In line 2 a global va\-ria\-ble is defined for the $\chi^2$ term, and in line 3 the number of threads is set, always equal to the number of devices. Each thread invokes the \texttt{Chi2} host function (Algorithm \ref{alg:chi2gpu}) with the appropriate channel data, in parallel. Once the function is done, a critical section guarantees exclusive access to workers to update the shared variable. Since the update consists of adding a scalar value to the global sum, we have decided to keep this computation on host.

Computation of $\nabla \chi^2$ follows the same logic as before (see Algorithm \ref{alg:dchi2openmp}). The major difference now is that the $\nabla \chi^2$ is a vector allocated on GPU number zero, and a kernel is used to sum up partial results. Recall that with P2P and UVA, kernels can read and write data from and to any GPU.

\begin{algorithm}
	\begin{algorithmic}[1]
		\STATE{\textbf{ParChi2}$(\vect{\widetilde{I}})$}
		\STATE{$\chi^{2} = 0$}
		\STATE{\textbf{set\_num\_threads}$(NDevices)$}
		\STATE{\textbf{\#pragma omp parallel for}}
		\FOR{$i=0$ to TOTALCHANNELS - 1}
		\STATE{\textbf{cudaSetDevice}$(i \% NDevices)$}
		\STATE{$\chi_i =  \text{\textbf{Chi2}}(\vect{\widetilde{I}}, \vect{V}^o_i, \vect{\omega}_i, x_c, y_c, \Delta u, \Delta v)$}
		\STATE{\textbf{\#pragma omp critical}}
		\STATE{$\chi^{2} = \chi^{2} + \chi^{2}_{i}$}
		\ENDFOR
	\end{algorithmic}
	\caption{Host multi-frequency $\chi^{2}$ function}
	\label{alg:chi2openmp}
\end{algorithm}

\begin{algorithm}
	\begin{algorithmic}[1]
		\STATE{\textbf{KParGradChi2}$(\vect{V}^R)$}
		\STATE{$\vect{\nabla \chi^{2}} = 0$}
		\STATE{\textbf{set\_num\_threads}$(NDevices)$}
		\STATE{\textbf{\#pragma omp parallel for}}
		\FOR{$i=0$ \TO TOTALCHANNELS - 1}
		\STATE{\textbf{cudaSetDevice}$(i \% NDevices)$}
		\STATE{$\vect{\nabla \chi^{2}}_i = \text{\textbf{KGradChi2}}(\vect{V}^R_i, \vect{\omega}_i, x_c, y_c, \Delta x, \Delta y)$}
		\STATE{\textbf{\#pragma omp critical}}
		\STATE{\textbf{KSumGradChi2}$(\vect{\nabla \chi^{2}}, \vect{\nabla \chi^{2}_{i}})$}
		\ENDFOR
	\end{algorithmic}
	\caption{Host multi-channel $\nabla \chi^{2}$ function}
	\label{alg:dchi2openmp}
\end{algorithm}

%% file: materials-methods.tex
% !TEX root = ./UVMEM_GPU_Paper.tex
% !BIB program = bibtex

\section{Experimental Settings}

In this section, we test our GPU version of MEM with three synthetic and three ALMA observatory data sets. 
We use them to demonstrate the effectiveness of the algorithm and to measure computational performance. As we pointed out in the introduction, this paper is not meant to be a comparison between MEM and CLEAN. This task would require an extensive imaging study which is beyond the scope of this work. Nevertheless, we find useful to display CLEAN images as a reference and because it is the most commonly method used today. 

\subsection{Data-sets}
\label{subsec:datasets}
A $1024\times 1024$ simulated image of a $7.5 \times 10^{-4}$ (Jy/beam) point source located at the image center and surrounded by a plateau was used to show that MEM can indeed reconstruct isolated point sources. Image pixel size was set to 0.003 arcsec and the plateau's magnitude was of $3.9 \times 10^{-4}$ (Jy/beam). Synthetic interferometer data was created with \texttt{ft} task from CASA, and the HL Tauri $uv$ coverage (see Table \ref{datasummary}). Image resolution was measured by the Full-Width-Half-maximum (FWHM) for MEM and CLEAN images. For this particular example, $\lambda= 0.1$ was used in MEM.

To further explore image resolution and other properties of MEM, we have created two simulated data sets using task \texttt{simobserve}. This task let us sample an input image using parameters such as antenna configurations, source direction, frequency, to name a few.

\begin{figure}[t]
    \centering
    \includegraphics[width=0.7\linewidth]{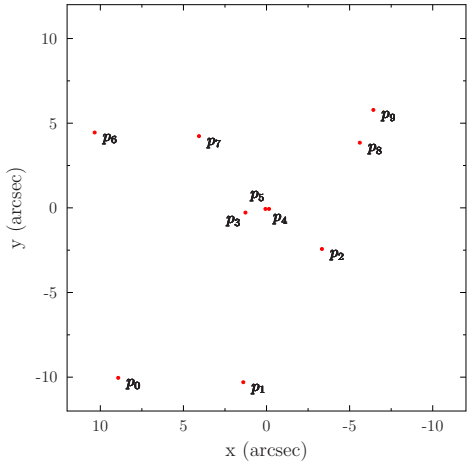}
    \caption{Point sources image.}
    \label{fig:ps_positions}
\end{figure}

Firstly, a $512\times512$ noiseless image of ten point sources distributed as shown in Figure \ref{fig:ps_positions} was simulated at 245 GHz. Point source intensities were randomly assigned normalized values between 0 and 1, and pixel size was set to 0.043 arcseconds. 
As it can be seen, point sources were located at random positions except for $p_4$ and $p_5$, which were placed 0.215 arcseconds apart. 

%\begin{figure}[h]
%\centering\includegraphics[width=0.7\linewidth]{Figure3C288Original}
%\caption{Model of radio galaxy 3C288.}
%\label{fig:3c288}
%\end{figure}

Secondly, a $256 \times 256$ image of a processed version of galaxy 3C288 \citep{radiogalaxy} with a pixel size set to 0.2 arcseconds, shown in Figure \ref{fig:3c288} was simulated at 84 GHz. The original image is an extended emission radio galaxy at 4.9 GHz with a mix of smooth, low resolution regions together with a small and high-resolution filament. Pre-processing implied clipping to zero all pixel values that were smaller than $10^{-3}$ (Jy/pixel), and adding a 1.0 Jy noise factor to simulated visibilities. Root mean square error (RMSE) of normalized values were measured in a circular region of interest shown in Figure \ref{fig:3c288}.

\input{Tabledatasummary}

We test our MEM implementation on real data summarized in Table \ref{datasummary}. The software CASA \citep{casa} was used to reconstruct CLEAN images that were not downloaded from the ALMA Science verification site and also to use the \texttt{mstransform} command to time average specific spectral windows of the HL Tau Band 6 dataset.

The first ALMA data set was the gravitationally lensed galaxy SDP 8.1 on Band 7 with four spectral windows, four channels and a total of 121,322,400 non-flagged visibilities \citep{sdp}. Channel 1 of spectral window 3 was completely flagged, so there were no visibilities from this particular channel. MEM images of $2048 \times 2048$ pixels were reconstructed, with $\lambda =0.0$. The CLEAN image  was downloaded from the ALMA Science Verification site, and had $3000 \times 3000$ pixels.

The second data set was the CO(6-5) emission line of the HD142527 protoplanetary disc, on band 9 with one channel of 107,494 visibilities \citep{co65}. This small data set was only used for code profiling and measuring GPU occupancy. Larger data sets could not be used to this end because profiling counters (32 bits) overflow with bigger reconstructions.

To test mosaic support the Antennae Galaxies Northern mosaic Band 7 data set was used. This data had 23 fields, 1 channel, and 149,390 visibilities. Once again, images of $512\times 512$ pixels were reconstructed with $\lambda = 0.1$ for MEM reconstruction.

Finally, the HL Tau Band 6 long baseline data set that corresponds to the observations of the young star HL Tauri surrounded by a protoplanetary disk \citep{hltau,hltau1,hltau2}. Spectral window zero (four channels) of this data set was used to measure speedup factor for varying image size. Data was time averaged on windows of 300 seconds, producing only 835,360 visibilities. To stress  multi-GPU compute capacity, a MEM image of $2048 \times 2048$ pixels was reconstructed from HL Tau full Band 6 with four spectral windows, four channels each and a total of 96,399,248 visibilities. The CLEAN image was downloaded from the ALMA Science Verification site and had $1600 \times 1600$ pixels. For reference, Table \ref{datasummary} lists relevant features of all ALMA data used.

\subsection{Computing performance metrics}
\label{subsec:performance}

Computational performance was measured by the speedup factor between one GPU and one CPU (single thread) wall-clock execution time versions of MEM, for three images sizes, namely $1024\times 1024$, $2048\times 2048$, and $4096\times 4096$. 
Since the number of MEM iterations depended on the data set, and some of the sequential reconstructions took excessively  long, the average execution time per iteration, was used as a timing measure. All speedup results were for short-spaced channels data sets. 
The CPU implementation is based on the conjugated gradient function \texttt{frprmn()} from the  Numerical Recipes book \citep{numericalrecipes}. It also employs the Fast Fourier transform, line search minimization, and line bracketing functions
from the same source. The GPU platform consisted of a cluster of four Tesla K80, each one with two Tesla GK210B GPUs, with 2496 streaming processors (CUDA cores) and 12 Gbytes of global me\-mo\-ry. It is important to highlight that this system had two PCI-Express ports on which two GPU were connected to each port. The CPU version was timed on an Intel Xeon E5-2640 V2 2.0 GHz processor.

Unfortunately, speedup cannot be measured in function of GPU cores. Speed factor is usually measured according to a certain number of processing elements like threads or processors. However, GPU programmers do not have control over the number of cores used in every kernel, but only over the grid dimensions. Grid size affects the streaming multiprocessors (SM) occupancy which is defined  as:

\begin{equation}
O = \frac{\text{Number of Active Warps}}{\text{Total Number of Warps}}
\end{equation}

Thus, occupancy measures how efficiently the SM are being used. Any program inefficiencies like inappropriate grid and block sizes, uncoalesced memory access, thread divergence or unbalanced workload will decrease the number of active warps and therefore SM occupancy and speedup. 
We used the NVIDIA Profiler \citep{cudaprofiler} to measure kernel's occupancy, floating-point operations, compute bound and memory bound kernels.  

\subsection{Image restoration} \label{sec:restoration}

Since MEM separates signal from noise \citep{libroAstro,libroAstro2}, the CASA package  was used to join the  MEM  model images and their residuals, following the technique of image restoration. Firstly,  the MEM model image was convolved with the clean elliptical beam, which effectively changes the MEM image units from Jy/pixel to Jy/beam, and also degrades its resolution. Secondly, a dirty image of each pointing was produced from the visibility residuals, using a user-specified weighting scheme \citep{ImagingBriggs}. These two last images were added to generate what is called a MEM restored image.

%% file: Tabledatasummary.tex
\begin{table*}
\centering
\caption{ALMA data sets summary.}
\label{datasummary}
\begin{tabular}{@{}rrcccr@{}}
\toprule
\textbf{Data set}  & \textbf{ADS/JAO.ALMA} & \textbf{Beam} &\textbf{SPWs} & \textbf{Channels/SPW} & \textbf{$Z$} \\
                   & \textbf{Code}         &                          &              &                       &  \\
\midrule
SDP 8.1 Band 7                  &        2011.0.000016.SV            &  $0.02'' \times 0.02''$    &  4                   &                4                         &            121,322,400              \\
HLTau Band 6                  &         2011.0.000015.SV          &    $0.04'' \times 0.02''$     &          4        &                  4                            &         96,399,248           \\

HLTau Band 6 SPW 0               &         2011.0.000015.S           &   $0.03'' \times 0.02''$    &            1        &                  4                            &         835,360         \\
Antennae North Band 7                         &       2011.0.00003.SV          & $1.06'' \times 0.68''$  &          1                 &              1                      &          149,390                   \\

HD142527 CO(6-5)  Band 9                        &      2011.0.00465.S          & $0.23'' \times 0.18''$   &        1                   &            1                   &       107,494                             \\
%3C288 Simulation at 84 GHz                     &      -         & $1.31'' \times 1.13''$   &        1                   &            1                   &       936,000                             \\
\bottomrule
\end{tabular}
\end{table*}

%% file: results.tex
% !TEX root = ./UVMEM_GPU_Paper.tex
% !BIB program = bibtex

\section{Results}
\label{subsec:imagingres}

\subsection{Computational performance}
\label{subsec:computationalperformance}

When using CPU timers, it is critical to remember that all kernel launches and API functions are asynchronous, in other words, they return control back to the calling CPU thread prior completing their work. Therefore, to accurately measured the elapsed time it is necessary to synchronize the CPU thread by calling \texttt{cudaDeviceSynchronize()} which blocks the calling CPU thread until all previously issued CUDA calls by the thread are completed \citep{cudabest}. 

%%
%% Table 1: Average time per iteration
%% This table appears in results but it is input here to edit the final look
\input{Tablecomputationalperformance}

Table \ref{tab:performanceperiteration} shows average time per iteration in mi\-nu\-tes, for the sequential and single GPU versions of MEM. Notice that for the HL Tau case, execution times increased proportionally with image size. Thus, going from a $1024\times 1024$ to a $4096\times 4096$ image, GPU execution time increased by a factor of 16, approximately. A similar be\-ha\-vior can be observed for the CPU case. However, the results for the CO(6-5) data sets are different, in this case the execution time and image size are not proportional. It must be remembered that this is the smallest data set used and neither CPU nor GPU are stressed to their maximal computational capacity. In any case, time differences are significant. A single iteration for the $4096\times 4096$ HL Tau sequential case took 38.2 days to finish, while the same reconstruction in GPU took only 9.85 minutes. Considering that the algorithm converged in 147 iterations in GPU, it is clear why we could not measure total execution time of the sequential version of MEM. Speedup factors for all cases are presented in  Table \ref{tab:performanceperiteration}. The smallest and largest speedup factors achieved were 1,638 and 5,579, respectively. The smallest speedup corresponded to the short baseline CO(6-5) with the smallest image size. Full reconstruction took 2.5 days in CPU and only 2.1 minutes in GPU.

% Table 2: Average speedup per iteration
%\input{Tablespeedup}

\begin{figure}
    \centering
    \includegraphics[width=0.6\linewidth]{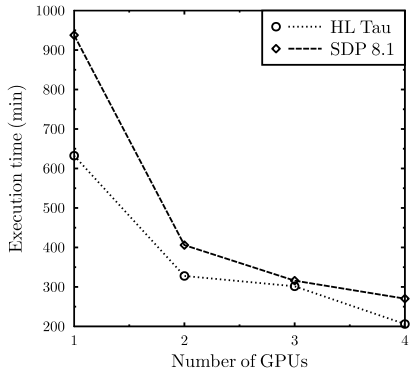}
    \caption{Average time per iteration using multi-GPU, SDP 8.1 Band 7 and HL Tau Band 6 data sets.}
    \label{fig:time_multigpu}
\end{figure}

Results for multi-GPU reconstruction are displayed in Figure \ref{fig:time_multigpu}. This graph shows average execution time per iteration as a function of number of GPUs, for the two largest data sets. The total number of visibilities of these data sets were similar, but the number of visibilities per channel were different. Also the SDP 8.1 data had one channel with zero visibilities.  Both curves display a typical fixed workload timing behavior, where execution time does not decrease linearly as more GPUs are used. Due to the fact that the number of visibilities in each channel varies, some load imbalance among GPUs may be causing this be\-ha\-vior. However, it is well known in parallel computing that fixed workload always has a maximum speedup possible.

For some of the most important kernels, Table \ref{tab:regperkernelperthread} lists number of registers per thread and achieved occupancy for the CO(6-5) data set.  
The \texttt{KGradChi2} kernel is the only kernel that achieved 100\% occupancy for any image size. This is probably due to the fact that it is the most compute intensive of all the kernels and it is able to maintain the GPU busy without many context-switches. Even though the other kernels require a smaller number of registers per thread than the \texttt{KGradChi2}, which means they can have more active warps per SM, they do not achieve maximum occupancy. Another observation which can be made is that occupancy varies differently for different kernels with image size. For instance, the \texttt{KGradPhi} kernel has a consistent 76\% to 78\% occupancy, but the \texttt{KReduce} kernel occupancy improves with image size, showing that the kernel itself is not big enough to fully exploit the GPU.

The excellent performance results can be explained by the fact that most implemented kernels do not move data to or from device.
Table \ref{tab:regperkernelperthread} also classifies each kernel according to a taxonomy proposed in \citep{Gregg2011}
based on memory-transfer overhead. 
As it can be seen all kernels except two are of the Non-Dependent (ND) type, which means they do not depend on data transfer between CPU and GPU. The \texttt{KAttenuation} kernel is a Single-Dependent Host-to-Device (SDH2D) type, because it requires
%data moved from host to GPU in order to proceed. But this is only true during the first iteration of the algorithm, after
data to be moved from host to GPU in order to proceed. But this is only true during the initialization step of the algorithm, after
which becomes a ND kernel for the rest of the iterations.
For instance, recall Algorithm \ref{alg:chi2} which is the host function to compute the $\chi^2$ term of the objective function. Once visibility data is on GPU, the
program proceeds as a pipeline of kernels which never transfer data back to host, until the last reduce kernel is
finished. Nevertheless, the data moved from device to host at the end of each iteration is just a scalar variable and only when the conjugate gradient achieves convergence, the final image and model visibilities are transferred to host.

%%
%% FIGURE 5: Average time per iteration multiGPU

%%
%% Table Occupancy
%%
\input{Tableoccupancy}

%%
%% Table Reconstructed point sources data
%%
%\input{TablepointsourcesFWHM}

\subsection{Image reconstruction}
\begin{figure}
    \centering
    \includegraphics[width=0.8\linewidth]{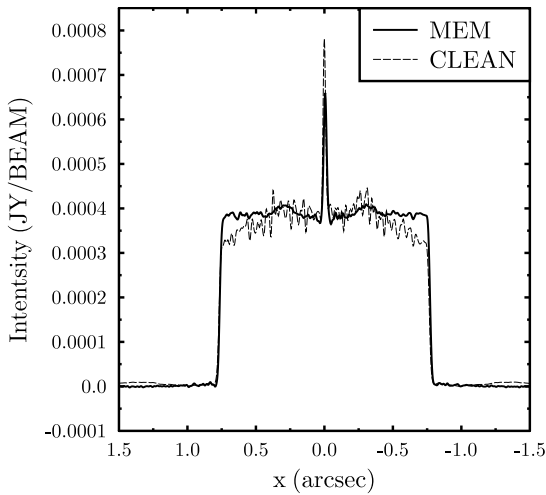}
    \caption{Reconstructed image profile of a point source in the center of a flat circular plateau, with MEM (restored), $\lambda=0.1$ and CLEAN. The point source and plateau intensities of the simulated image are $7.5 \times 10^{-4}$ (Jy/beam) and $3.9\times 10^{-4}$ (Jy/beam), respectively.}
    \label{fig:plateau_mem_restored}
\end{figure}

Figure \ref{fig:plateau_mem_restored} depicts CLEAN and restored MEM image profiles of the isolated point source image. On one hand, CLEAN achieved a higher point source intensity, but on the other hand MEM performed a smoother fit of the plateau. The FWHM was 0.064 and 0.062 arcsec for MEM and CLEAN, respectively. Notice that CLEAN overestimated the point source intensity and MEM underestimated the value. CLEAN tends to fit the point source and plateau overestimating the peak, but MEM is known to lose flux in recovered images \citep{MEM}.

For the field of point sources (Figure \ref{fig:ps_positions}), CLEAN delivered uniform resolution across the entire field of view (0.142 arcsecs) and even though MEM ($\lambda = 0.005$ ) produced better resolution than CLEAN, values greatly depend on image location. For instance, FWHM for $p_0$ and $p_9$ were 0.09 and 0.07 arcseconds, respectively, but for point sources near the center like  $p_3$ and $p_2$, FWHM was 0.05.
Since the distance between the two center points $p_4$ and $p_5$ was 0.215 arcsec, resolution at these locations could no be computed in the CLEAN image as there are only 1.5 pixels to fit a Gaussian. MEM instead delivered a FWHM of 0.05 arcseconds for both points.
For intensity reconstruction, CLEAN was able to
recover over 99.7\% of the signal strength in all cases, and MEM recovered differently at different locations. For point sources away from the center, like $p_0$ and $p_1$ peak recovery was 98\% and 99\%, respectively, but for point sources near the center, like $p_2$ and $p_3$, recovery was only 85\% and 74\%, respectively.  

\begin{figure*}
\centering
    \subfloat[Model]{
    \includegraphics[width=0.3\linewidth]{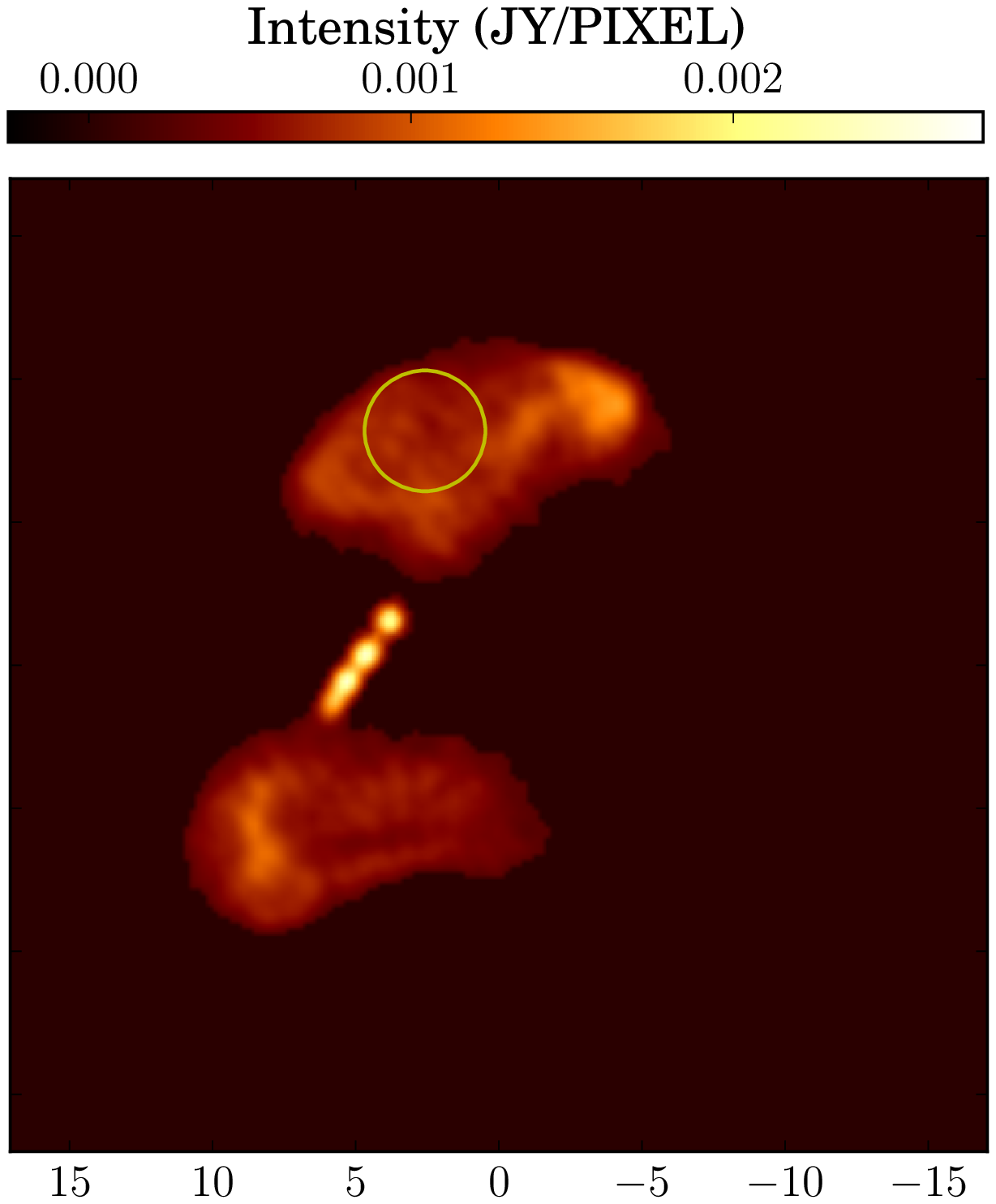}
    \label{fig:3c288}
    }
    \subfloat[CLEAN]{
    \includegraphics[width=0.3\linewidth]{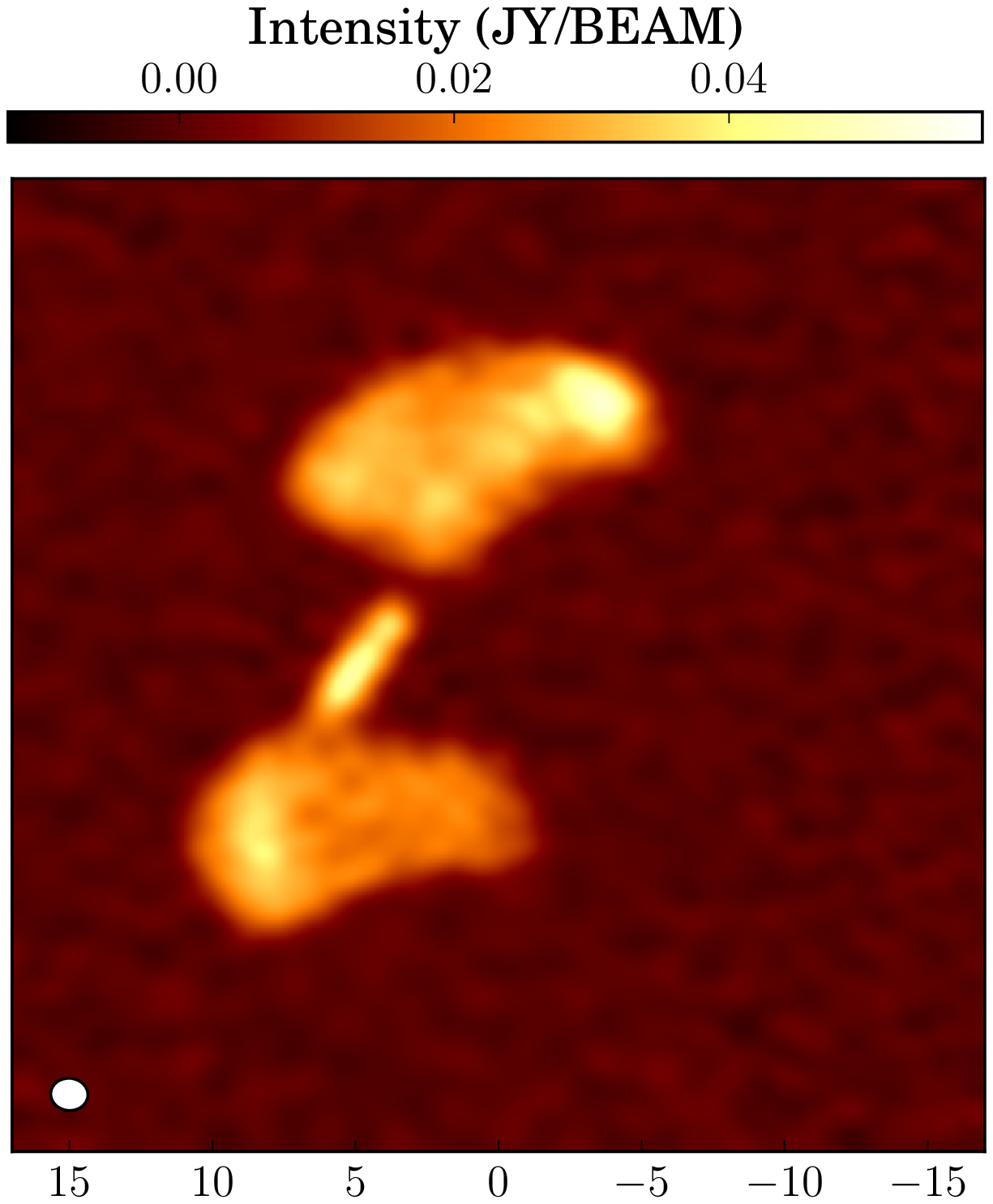}
    \label{subfig:3c288-clean}
    }
    \subfloat[MEM ($\lambda = 0.01$)]{
    \includegraphics[width=0.3\linewidth]{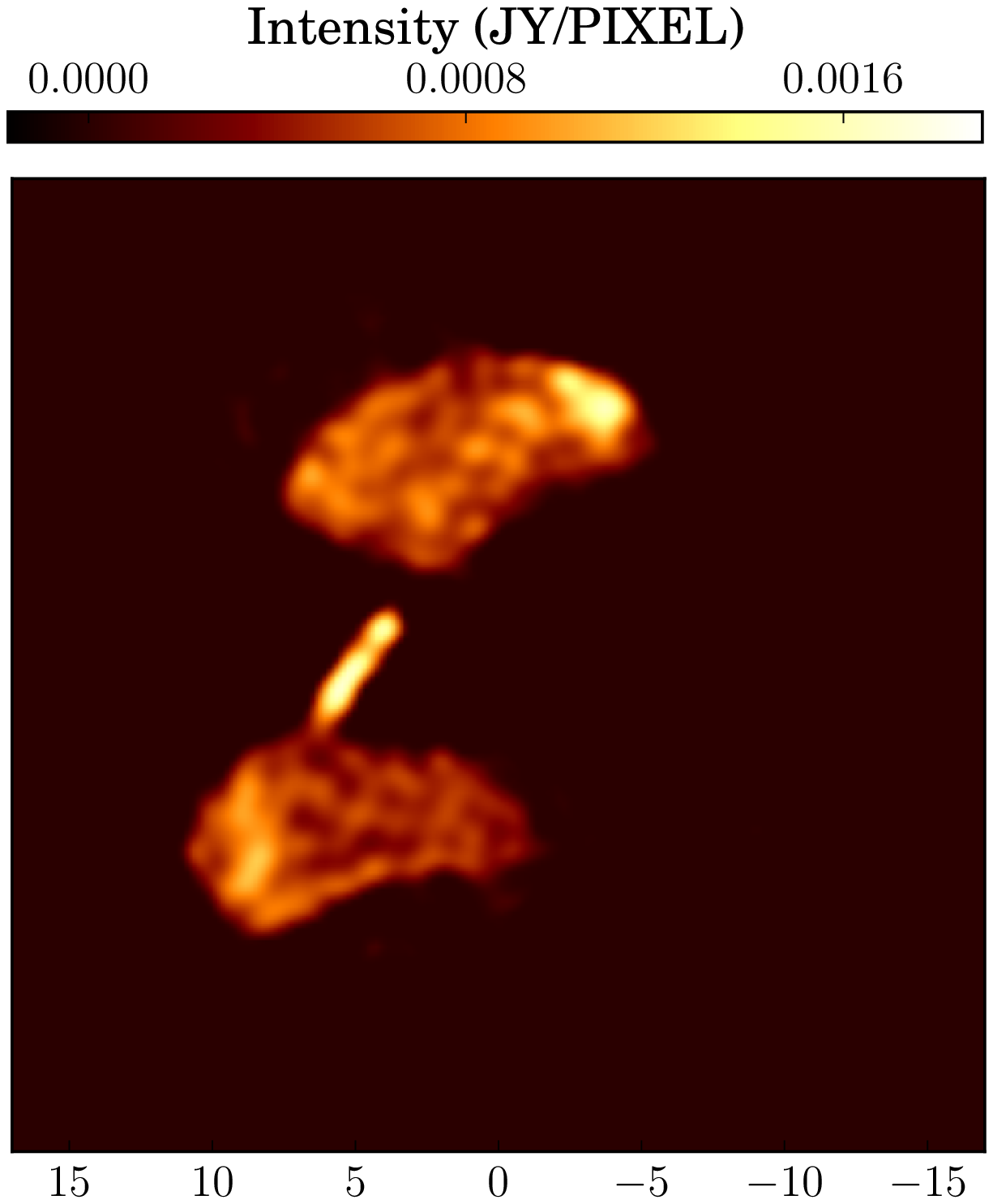}
    \label{subfig:3c288-mem}
    }

\caption{(a) Model of a modified version of radio galaxy 3C288, (b) CLEAN image (with a beam of $1.31'' \times 1.13''$) and (c) MEM at 84 GHz.}
\label{fig:extendedresults}
\end{figure*}

Images for the last simulated data set are displayed in Figures \ref{subfig:3c288-clean} and \ref{subfig:3c288-mem}. It is clearly noticeable that MEM does a better job removing systematic background noise. In particular at the region of interest, RMSE of normalized reconstructions reach 0.0239 for CLEAN versus a closer 0.0159 for MEM. Filament in the middle of the image is better resolved in MEM due to its smaller resolution. This corresponds the same phenomena found previously in the point source field. It is worth mentioning that CLEAN images for the first two data sets were reconstructed without user supervision, but for the extended emission example, there were necessary 10 interactive cycles of 1000 iterations each.

\begin{figure*}
    \centering
    \subfloat[]{
    \includegraphics[width=0.3\linewidth]{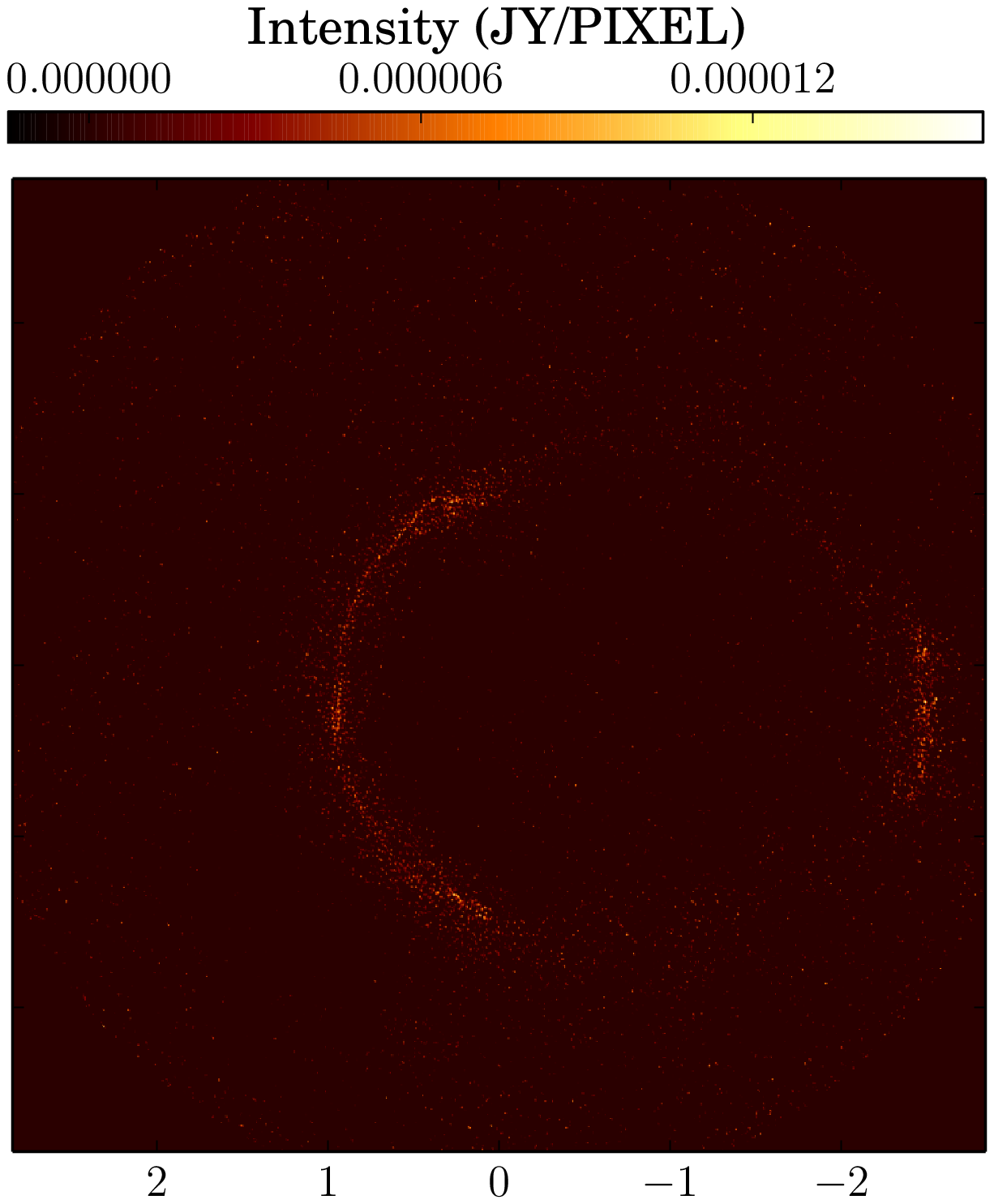}
    \label{subfig:sdpb7-model}
    }
    \subfloat[]{
    \includegraphics[width=0.3\linewidth]{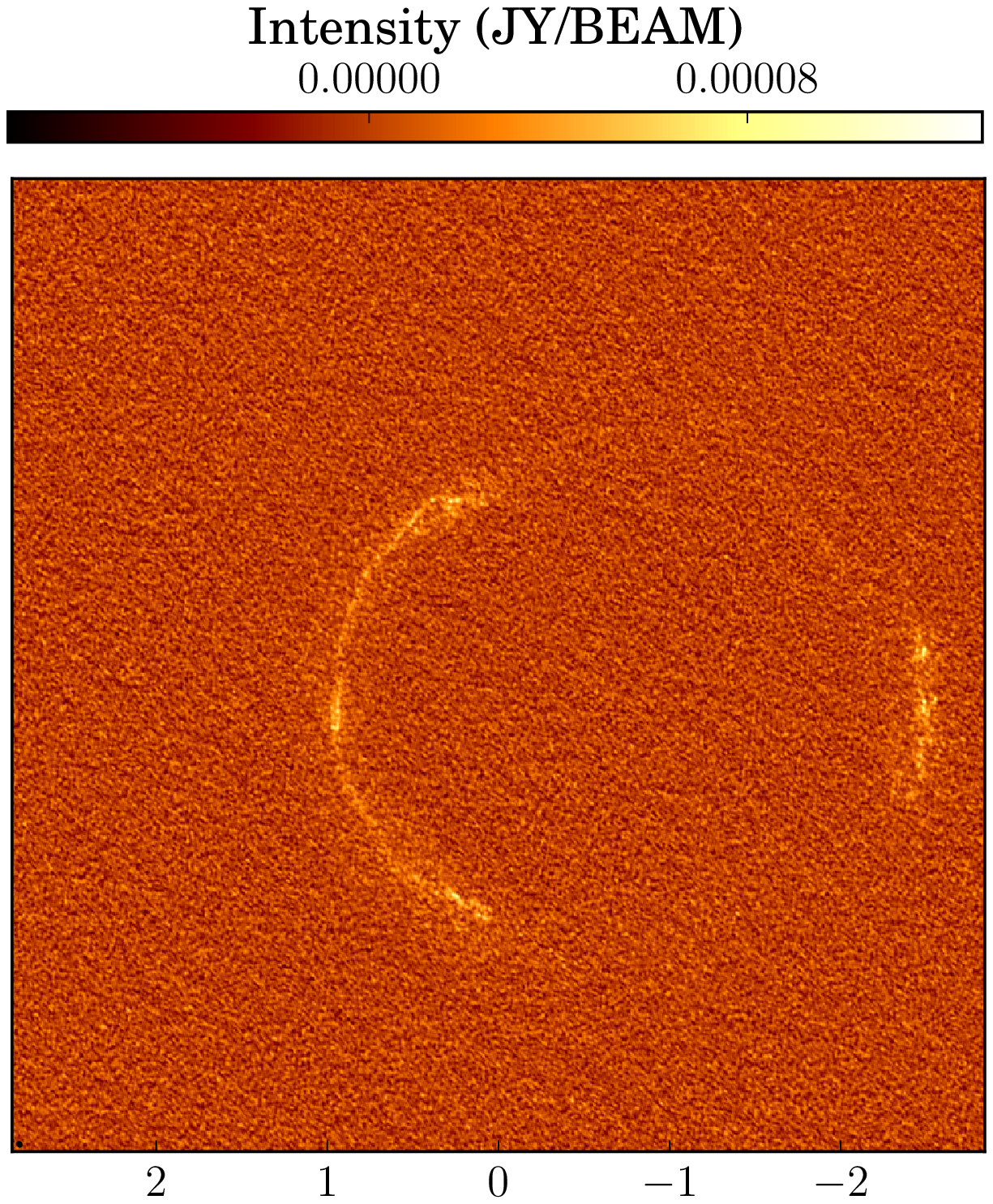}
    \label{subfig:sdpb7-restored}
    }
    \subfloat[]{
    \includegraphics[width=0.3\linewidth]{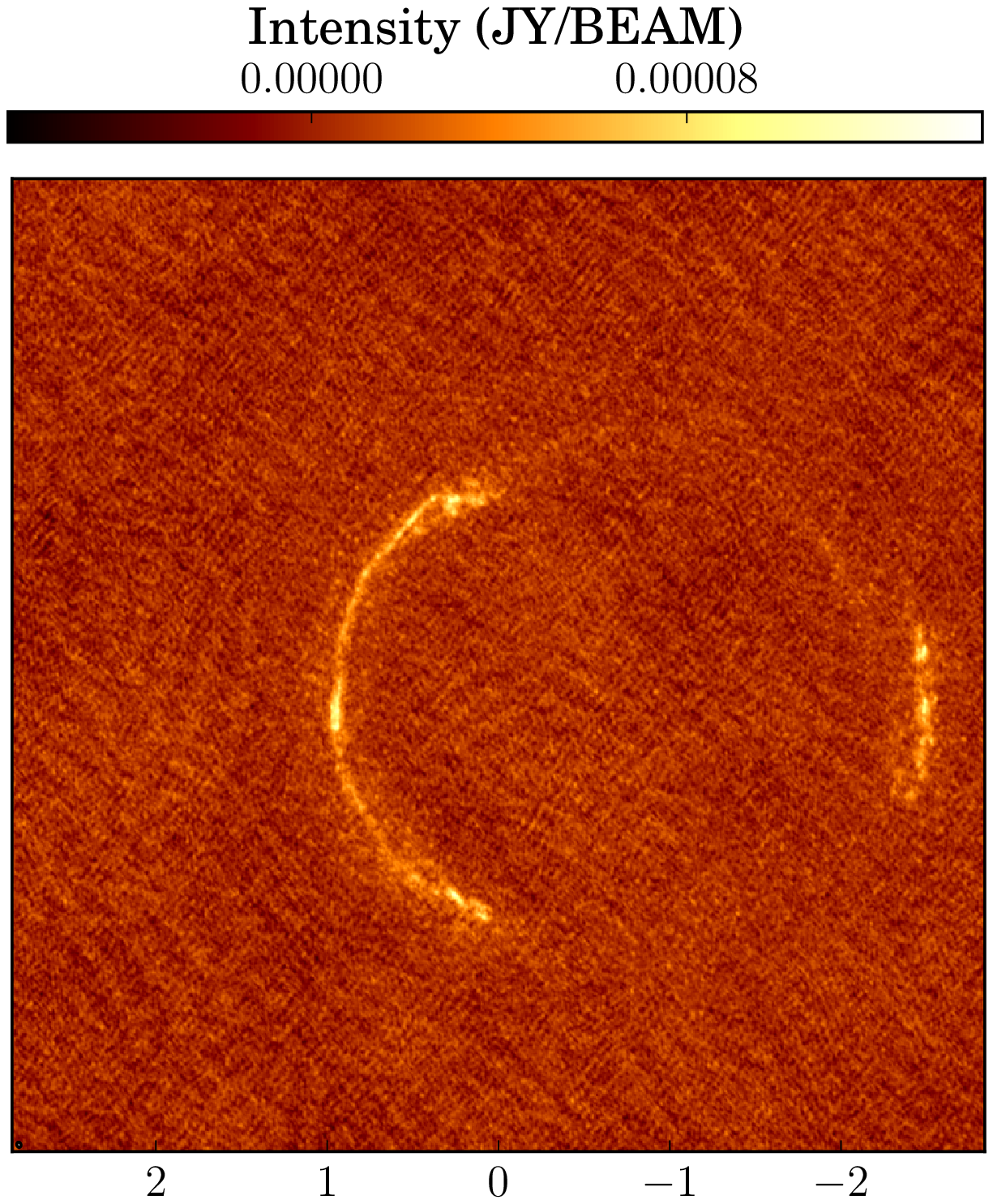}
    \label{subfig:sdpb7-clean}
    }
    \\
    \subfloat[]{
    \includegraphics[width=0.3\linewidth]{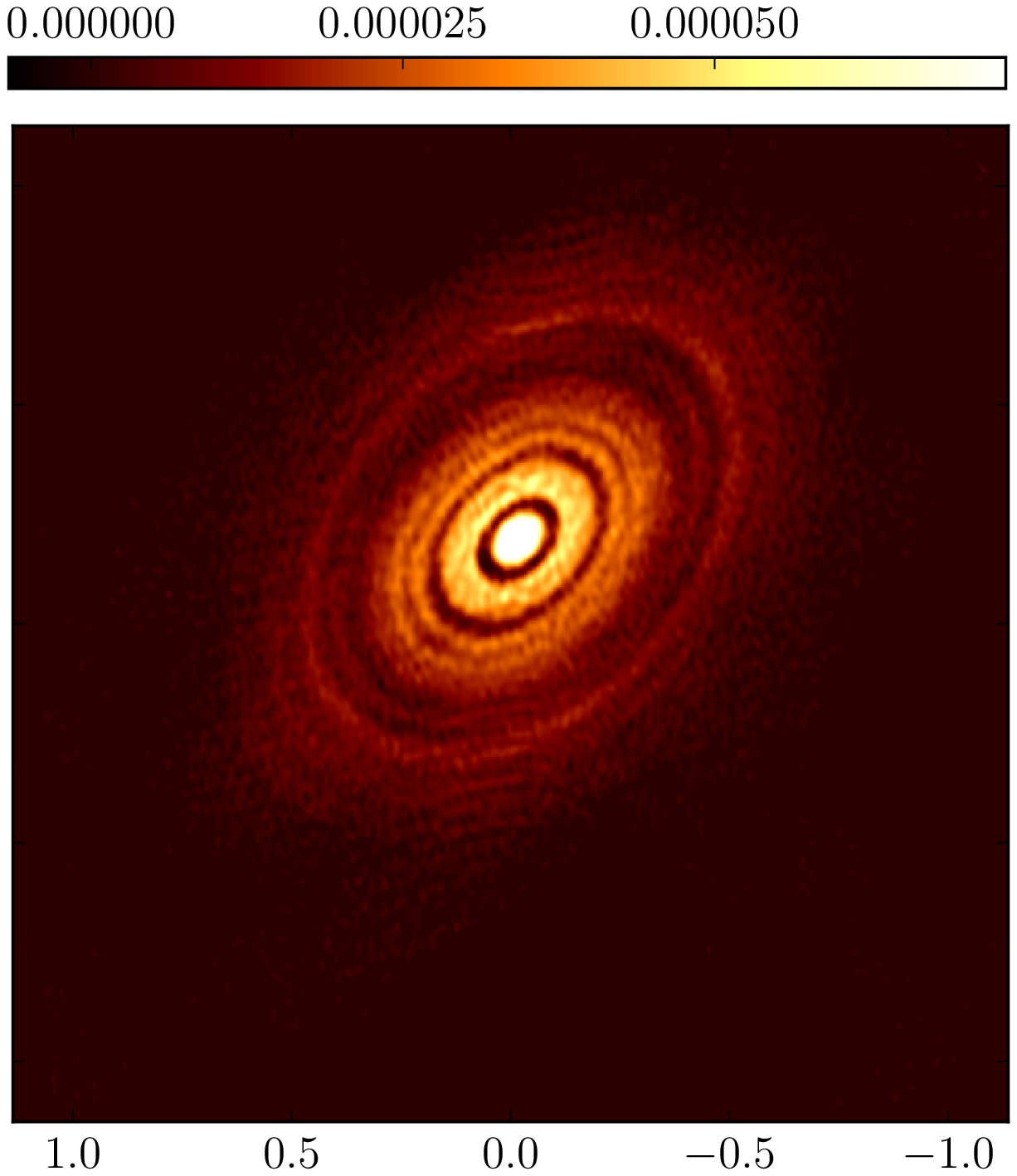}
    \label{subfig:hltau-model}
    }
    \subfloat[]{
    \includegraphics[width=0.3\linewidth]{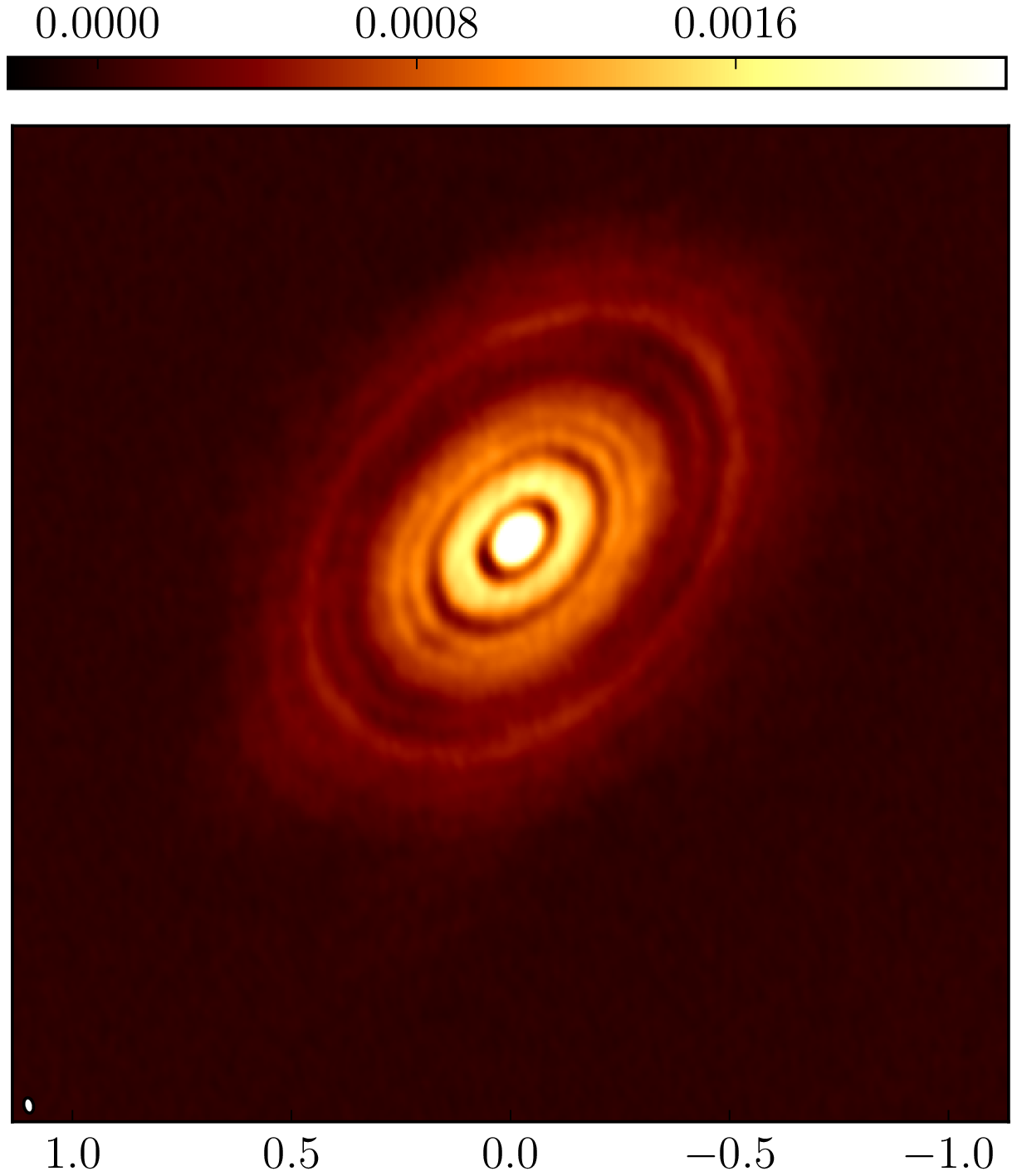}
    \label{subfig:hltau-restored}
    }
    \subfloat[]{
    \includegraphics[width=0.3\linewidth]{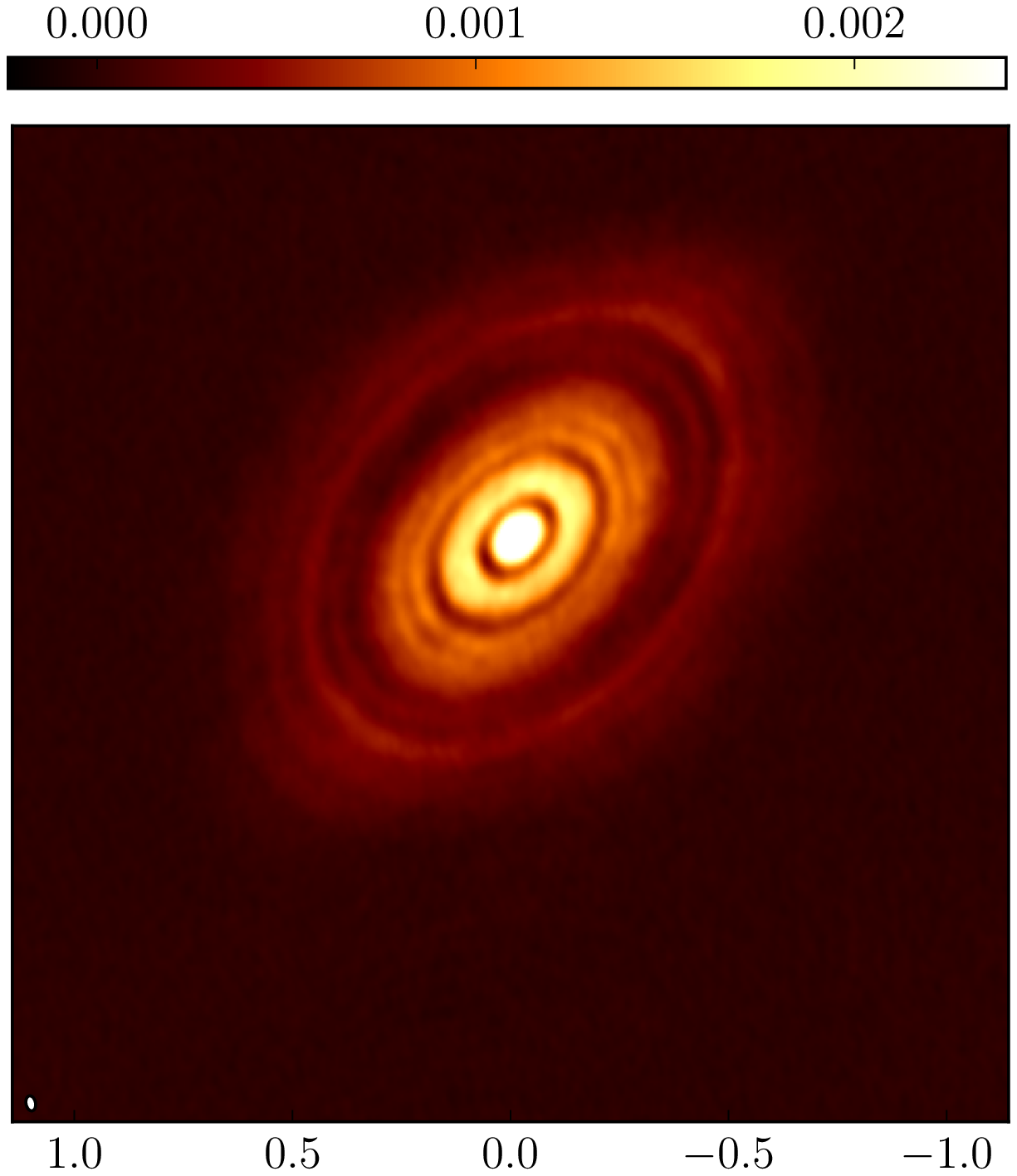}
    \label{subfig:hltau-clean}
    }
    \label{fig:hltauresults}    
	\\
	\subfloat[]{
    \includegraphics[width=0.3\linewidth]{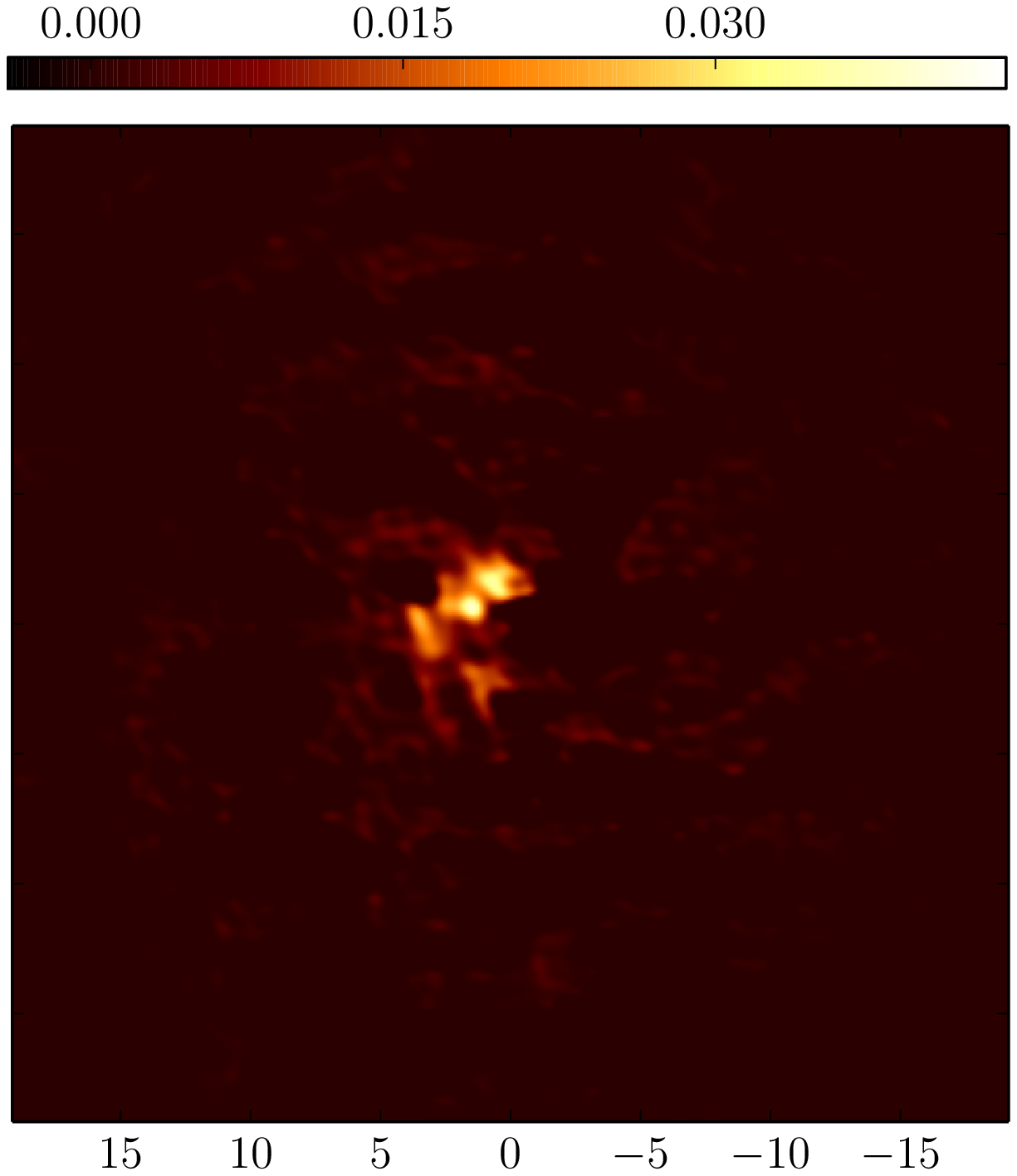}
    \label{subfig:antennae-model}
    }
    \subfloat[]{
    \includegraphics[width=0.3\linewidth]{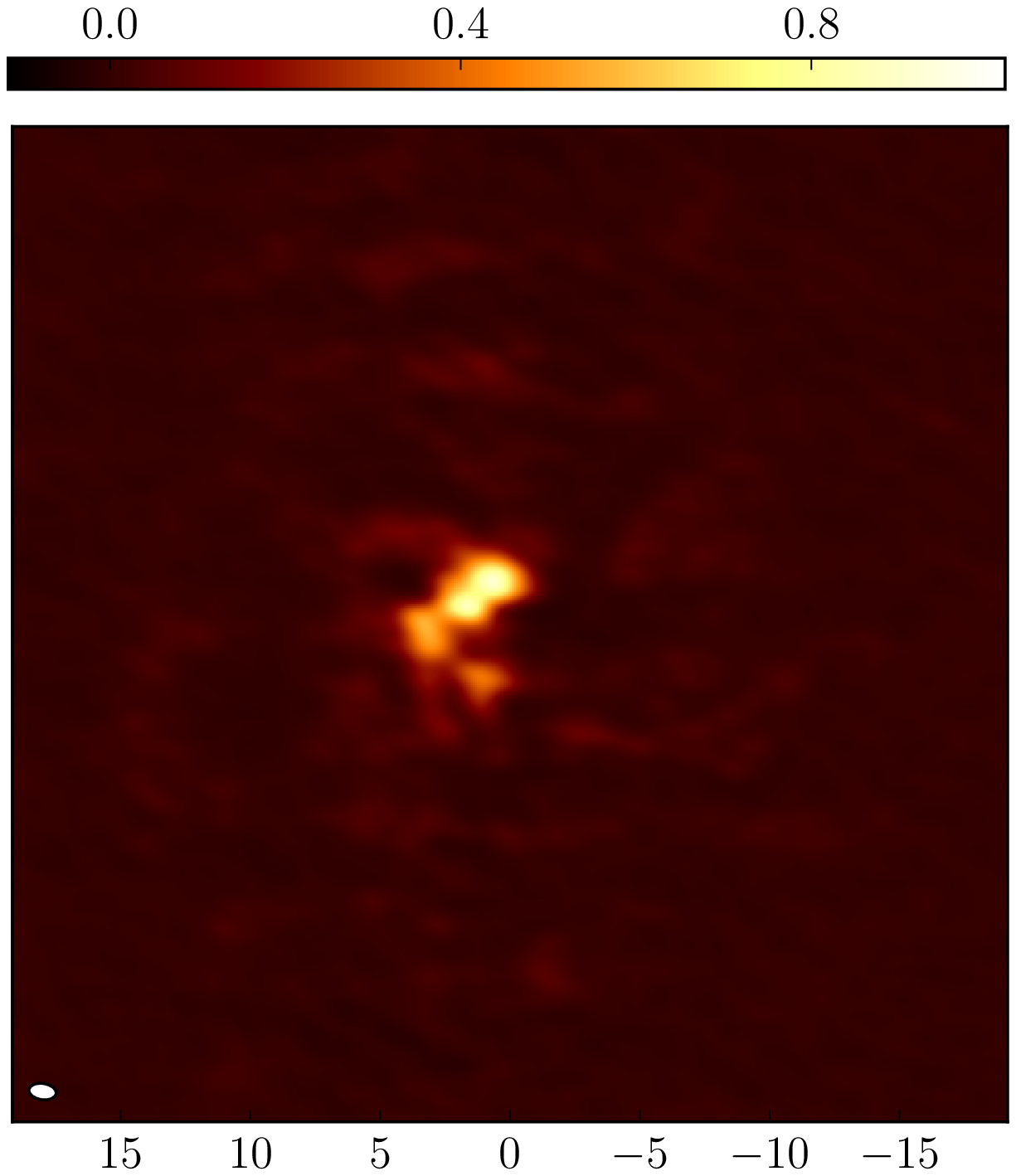}
    \label{subfig:antennae-restored}
    }
    \subfloat[]{
    \includegraphics[width=0.3\linewidth]{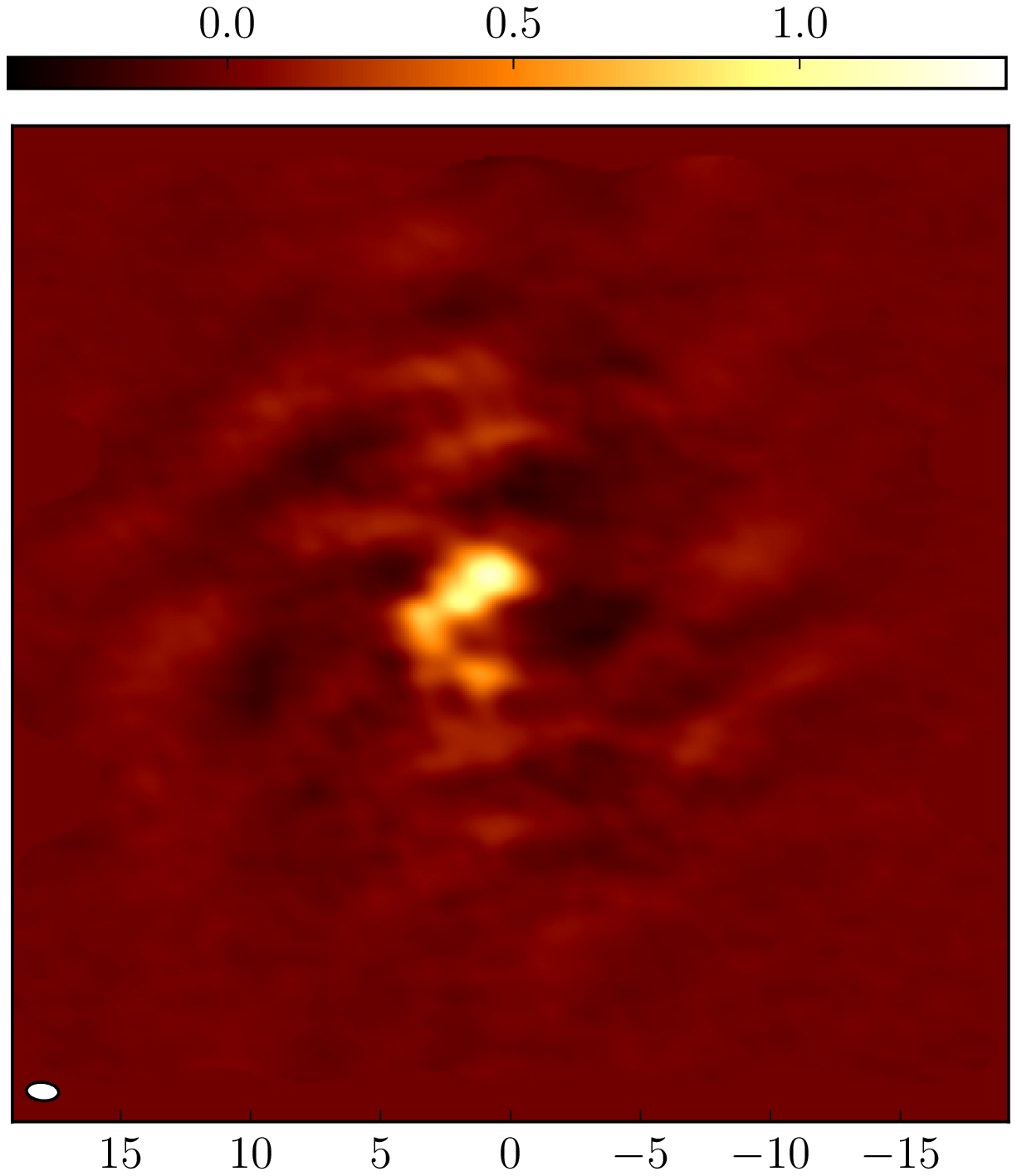}
    \label{subfig:antennae-clean}
    }	
	\caption{Image synthesis results using MEM and CLEAN. First column shows MEM model images. Second column shows MEM Restored images, and third column shows CLEAN images. (a)-(c) SDP 8.1 on Band 7, (d)-(f) HL Tau on Band 6. (g)-(i) Antennae Galaxies Northern mosaic on Band 7.}
	\label{fig:imagematrix}
\end{figure*}

Image reconstruction results for real data are shown in Figure \ref{fig:imagematrix}. First column displays MEM model images, second co\-lumn shows MEM restored images, and third column shows CLEAN images. First and second rows corresponds to the SDP 8.1 on Band 7 and HL Tau on Band 6, respectively. As has been said, both are long baseline data sets, and therefore high angular resolution and greater detail are expected. Third row shows the Antennae Galaxies North on Band 7. This dataset has short baselines and is also a mosaic. Model images (first column) show images with only signal as MEM separates it from noise, and in all cases restored images are of lower resolution than model images. 

%% file: Tablecomputationalperformance.tex
% !TEX root = ./UVMEM_GPU_Paper.tex
% !BIB program = bibtex

%\begin{table*}
%\centering
%\caption{Average time per iteration (min) for one CPU and a single GPU versions of MEM, for two data sets and different images sizes.}
%\label{tab:timesperiteration}
%\begin{tabular}{@{}lrrrrrrr@{}}
%\toprule
%\multicolumn{1}{c}{\textbf{Data}} & \multicolumn{3}{c}{\textbf{CPU time}}  & \multicolumn{1}{c}{} & \multicolumn{3}{c}{\textbf{GPU time}}  \\ %\midrule 
%                                  & 1024x1024  & 2048x2048   & 4096x4096   &                      & 1024x1024 & 2048x2048 & 4096x4096 \\
%\cmidrule(lr){2-4}
%\cmidrule(l){6-8}
%CO(6-5)                           & $131$  & $349$   & $360$   &                      & $0.08$   & $0.11$   & $0.13$   \\
%HL Tau B6w0                       & $2,645$ & $10,339$ & $55,009$ &                      & $0.60$   & $2.39$   & $9.86$   \\ \bottomrule
%\end{tabular}
%\end{table*}

\begin{table*}
\centering
\caption{Average time per iteration (min) and speed factor for single CPU and single GPU versions of MEM,
for two data sets and varying image size.}
\label{tab:performanceperiteration}
\begin{tabular}{@{}rrrrr@{}}
\toprule
\textbf{Image Size}        & \textbf{Data set}   & \textbf{CPU time}  & \textbf{GPU time} & \textbf{Speedup}  \\ \midrule 
%\multirow{2}{*}{$1024\times 1024$} & CO(6-5)    & $131$     & $0.08$ & $1,638$ \\ 
$1024\times 1024$          & CO(6-5)    & $131$     & $0.08$ & $1,638$ \\ 
                           & HLTau B6w0 & $2,645$   & $0.60$ & $4,408$ \\ 
                           \cline{2-5}
$2048\times 2048$          & CO(6-5)    & $349$     & $0.11$ & $3,173$ \\ 
                           & HLTau B6w0 & $10,339$  & $2.39$ & $4,326$ \\ 
                           \cline{2-5}
$4096\times 4096$          & CO(6-5)    & $360$     & $0.13$ &  $2,769$ \\ 
                           & HLTau B6w0 & $55,009$  & $9.86$ & $5,579$ \\  \bottomrule
\end{tabular}
\end{table*}

%% file: Tableoccupancy.tex
% !TEX root = ./UVMEM_GPU_Paper.tex
% !BIB program = bibtex

\begin{table*}
\centering
\caption{Occupancy and taxonomy \citep{Gregg2011} of the most important kernels using a 1024 thread block size and different image sizes.}
\label{tab:regperkernelperthread}
\begin{tabular}{@{}lrcccc@{}}
\toprule
\multirow{2}{*}{\textbf{Kernel}} & \multicolumn{1}{l}{\multirow{2}{*}{\textbf{Taxonomy}}} & \multicolumn{1}{l}{\multirow{2}{*}{\textbf{Registers per thread}}} & \multicolumn{3}{c}{\textbf{Achieved Occupancy (\%)}} \\ \cmidrule(l){4-6} 
                                 & \multicolumn{1}{l}{}                                   & \multicolumn{1}{l}{}                                               & 1024x1024        & 2048x2048       & 4096x4096       \\ \cmidrule(r){1-3} \cmidrule(l){4-6}
KGradPhi                         & ND                                                     & 10                                                                 & 77               & 76              & 78              \\
KChi2Res                         & ND                                                     & 12                                                                 & 81               & 79              & 80              \\
KGradChi2                        & ND                                                     & 30                                                                 & 100              & 100             & 100             \\
KEntropy                         & ND                                                     & 13                                                                 & 86               & 81              & 79              \\
KGradEntropy                     & ND                                                     & 13                                                                 & 87               & 81              & 80              \\
KAttenuation                     & SDH2D/ND                                                     & 24                                                      & 89               & 90              & 90              \\
KInterpolation                   & ND                                                     & 32                                                                 & 92               & 90              & 88              \\
KModulation                      & ND                                                     & 18                                                                 & 89               & 89              & 88              \\
KReduce                          & SDD2H                                                   & 13                                                                 & 75               & 88              & 96              \\ \bottomrule
\end{tabular}
\end{table*}

%% file: conclusion.tex
% !TEX root = ./UVMEM_GPU_Paper.tex
% !BIB program = bibtex

\section{Conclusions}
We have developed a high performance computing solution for maximum entropy image reconstruction in interferometry. The solution is based on GPU for single channel data and on multiple GPUs for multi-spectral data. The implementation uses a host algorithm that runs the main iteration loop and orchestrates kernel calls. We have decided to write small kernels to improve data locality and minimize thread divergence. Most data is kept on device memory which also minimizes memory moves between host and devices. Overall, we have found that the algorithm renders naturally into a SIMT paradigm and makes it a good candidate for successful GPU implementation. 

The resulting code achieves a speedup of approximately 1681 times for our smallest data set, which means that CO(6-5) can be reconstructed in 2.1 minutes, instead of 2.5 days in single CPU. Long baseline HL Tau Band 6, spectral window zero and $1024^2$ image size, takes 58 minutes approximately. However for full band,  multi-spectral data sets, like SDP 8.1 and HL Tau, non-gridded MEM reconstruction is still a challenging task. For instance, full data-set band 6 HL Tau reconstruction requires 200 minutes per iteration, which is still expensive for routine use with the considered hardware.

Results of image synthesis with MEM has demonstrated that the method removes systematic background noise, has a better resolution, but reaches less intensities at point sources. Moreover, MEM has demonstrate to have less signal variations in flat areas, and having smoother fit than CLEAN on extended structures. This could clearly benefit MEM users by allowing them to identify a finer morphology of the objects, but at the cost of losing signal intensity. 

We would like to emphasize that although this work implements the MEM algorithm using the conjugate gradient algorithm, many other first order optimization methods based on gradient can be used. Furthermore, the design permits the use of other well behaved regularization terms to penalize the solution at a similar cost. In consequence we demonstrate that non-linear optimization methods implemented in GPU can be used in practice for research in image synthesis. 

Finally, each image synthesis method has its own bias, but making use of many of them could improve the astrophysical analysis of reconstructions. Therefore, a proven high performance MEM implementation could contribute as a tool for image assessment.

The code used in this work is licensed under the GNU General Public License (GNU GPL) v3.0 and is freely available at \url{https://github.com/miguelcarcamov/gpuvmem}.

\section{Future work}

Results are encouraging, but there are still several cha\-llen\-ges that need to be addressed. Most of the algorithm's time is spent on gradient computation, and for nearly-black objects, evaluation on background pixels far from the source 
carry little or zero contribution to the signal. Therefore, we are currently modifying our code to include a focus-of-attention improvement which will allow the  user to select a region on the image where reconstruction should take place, thus reducing substantially the number of free parameters. We are also working on a solution to reconstruct data sets with long-spaced multiple channels, also called multi-frequency synthesis. 
Finally, we are now in a position to conduct a throughout experimental evaluation of MEM imaging parameters including resolution, signal-to-noise ratio, signal recovery, and associated uncertainties for real and simulated data sets. 

%% file: appendix.tex
\section{Beam size}
\label{apx:beam}
Provided some degree of $uv$-coverage, a zero-order approximation to the sky signal can be obtained from the  dirty image
\begin{align}
\label{eq:dirty}
I^D(X) = \sum_k \omega_k V^G_k \exp(2\pi i U_k \cdot X), \\
\notag ~\text{with}, \omega_k = (1/\sigma_k^2) / \sum_l (1/\sigma_l^2).
\end{align}
where $V^G_k$ are the gridded visibilities \citep{ImagingBriggs}.

%where $U_k$ are the sampling locations in Fourier space.

$I^D(X)$ is called the {\em natural-weights} dirty map and has units of Jy~beam$^{-1}$, where the beam is the solid angle subtended by the FWHM of the Point Spread Function (PSF)  
\begin{align}
    \label{eq:PSF} \text{PSF}(X) = \sum_k \omega_k \cos{(U_k \cdot X)}
\end{align}
The beam is usually approximated by fitting an elliptical Gau\-ssian on the PSF shape and considering the FWHM. The parameters of the ellipse are BMAJ: Major diameter, BMIN: Minor diameter, and BPA: anti-counter clockwise angle in degree. This shape is interpreted as a natural resolution for the CLEAN algorithm.
\section{Thermal noise}
\label{apx:thermal}
Under the assumption of uncorrelated samples\footnote{visibility data on $uv$ points closer than the size of the dishes are correlated} we can obtain an analytic expression for the thermal noise $\sigma_D$ using the following error propagation from dirty image $I^D$
\begin{align}
    \notag \sigma_D^2 &=\mathrm{Var}(I^D)=\mathrm{Var}\biggl(\text{Re}\biggl(\sum_k{\frac{\omega_k}{\sum_l{\omega_l}}\bigl(V^o_k e^{-2\pi i \theta_k}\bigr)}\biggr)\biggr)=\\
    \notag &=\text{Var}\biggl(\sum_k{\frac{\omega_k}{\sum_l{\omega_l}}\bigl(V^{oR}_k\cos(\theta_k)-V^{oI}_k\sin(\theta_k)\bigr)}\biggr)\\
    \notag &= \sum_k{\frac{\omega_k^2}{(\sum_l{\omega_l})^2}\biggl(\frac{1}{\omega_k}\cos^2(\theta_k)+\frac{1}{\omega_k}\sin^2(\theta_k)\biggr)}\\
    \notag &= \frac{\sum_k{\omega_k}}{(\sum_k{\omega_k})^2} \\
    \notag \sigma_D &= \frac{1}{\sqrt{\sum_k{\omega_k}}}=\frac{1}{\sqrt{\sum_k {\frac{1}{\sigma^2_k}}}}
\end{align}
where $\theta_k=U_k\cdot X$, $\omega_k=\frac{1}{\sigma^2_k}$. We also assume uncorrelated imaginary and real part of the samples. $\sigma_D$ units are Jy~beam$^{-1}$. Then, the constant to change units from Jy~beam$^{-1}$  to Jy~pixel$^{-1}$ is 
\begin{align}
    \notag \frac{(\Delta x)^2}{(\pi/(4*\log(2)) * \text{BMAJ} * \text{BMIN}}  
\end{align}
    where $\Delta x$ is the pixel side size in radians.
\section{$\chi^2$ gradient}
\label{apx:chi2gradient}
The gradient of a model visibility is approximated from the DFT of model image.
\begin{align}
    \notag \frac{\partial V^R_k}{\partial I_l} = \frac{\partial V^m_k}{\partial I_l} \sim \frac{\partial}{\partial I_l} \sum_j{I_j e^{-2\pi i \theta_{kj}}} = e^{-2\pi i \theta_{kl}} 
\end{align}
Where $\theta_{kl}=U_k\cdot X_l$, from previous approximation we obtain
\begin{align}
    \notag \frac{\partial}{\partial I_l} \chi^2 &= \frac{\partial}{\partial I_l} \frac{1}{2}\sum_k{\frac{\overline{V^R_k}V^R_k}{\sigma_k^2}} \\
    \notag &=\frac{1}{2}\sum_k{\frac{1}{\sigma_k^2}\biggl(\frac{\partial \overline{V^R_k}}{\partial I_l}V^R_k+\overline{V^R_k}\frac{\partial V^R_k}{\partial I_l}\biggr)}  \\
    \notag &= \sum_k{\frac{1}{\sigma_k^2}\text{Re}\biggl(\frac{\partial \overline{V^R_k}}{\partial I_l}V^R_k\biggr)} = \sum_k{\frac{\text{Re}\bigl(V^R_k e^{2\pi i \theta_{kl}}\bigr)}{\sigma^2_k}}
\end{align}

%% file: UVMEM_GPU_Paper.bbl
\begin{thebibliography}{46}
\expandafter\ifx\csname natexlab\endcsname\relax\def\natexlab#1{#1}\fi
\providecommand{\url}[1]{\texttt{#1}}
\providecommand{\href}[2]{#2}
\providecommand{\path}[1]{#1}
\providecommand{\DOIprefix}{doi:}
\providecommand{\ArXivprefix}{arXiv:}
\providecommand{\URLprefix}{URL: }
\providecommand{\Pubmedprefix}{pmid:}
\providecommand{\doi}[1]{\href{http://dx.doi.org/#1}{\path{#1}}}
\providecommand{\Pubmed}[1]{\href{pmid:#1}{\path{#1}}}
\providecommand{\bibinfo}[2]{#2}
\ifx\xfnm\relax \def\xfnm[#1]{\unskip,\space#1}\fi
%Type = Article
\bibitem[{{ALMA Partnership} et~al.(2015)}]{hltau}
\bibinfo{author}{{ALMA Partnership}}, et~al., \bibinfo{year}{2015}.
\newblock \bibinfo{title}{{The 2014 ALMA Long Baseline Campaign: First Results
  from High Angular Resolution Observations toward the HL Tau Region}}.
\newblock \bibinfo{journal}{Astrophysical Journal Letters}
  \bibinfo{volume}{808}, \bibinfo{pages}{L3}.
%Type = Article
\bibitem[{{Bridle} et~al.(1989){Bridle}, {Fomalont}, {Byrd} and
  {Valtonen}}]{radiogalaxy}
\bibinfo{author}{{Bridle}, A.H.}, \bibinfo{author}{{Fomalont}, E.B.},
  \bibinfo{author}{{Byrd}, G.G.}, \bibinfo{author}{{Valtonen}, M.J.},
  \bibinfo{year}{1989}.
\newblock \bibinfo{title}{{The unusual radio galaxy 3C 288}}.
\newblock \bibinfo{journal}{The Astrophysical Journal} \bibinfo{volume}{97},
  \bibinfo{pages}{674--685}.
%Type = Inproceedings
\bibitem[{Briggs et~al.(1999)Briggs, Schwab and Sramek}]{ImagingBriggs}
\bibinfo{author}{Briggs, D.S.}, \bibinfo{author}{Schwab, F.R.},
  \bibinfo{author}{Sramek, R.A.}, \bibinfo{year}{1999}.
\newblock \bibinfo{title}{Imaging}, in: \bibinfo{editor}{Taylor, G.B.},
  \bibinfo{editor}{Carilli, C.L.}, \bibinfo{editor}{Perley, R.A.} (Eds.),
  \bibinfo{booktitle}{Synthesis Imaging in Radio Astronomy II}, pp.
  \bibinfo{pages}{127--149}.
%Type = Book
\bibitem[{Burger and Burge(2010)}]{imageProcessing}
\bibinfo{author}{Burger, W.}, \bibinfo{author}{Burge, M.},
  \bibinfo{year}{2010}.
\newblock \bibinfo{title}{Principles of Digital Image Processing: Core
  Algorithms}.
\newblock Undergraduate Topics in Computer Science,
  \bibinfo{publisher}{Springer London}.
%Type = Article
\bibitem[{Cabrera et~al.(2008)Cabrera, Casassus and
  Hitschfeld}]{VoronoiCabrera}
\bibinfo{author}{Cabrera, G.F.}, \bibinfo{author}{Casassus, S.},
  \bibinfo{author}{Hitschfeld, N.}, \bibinfo{year}{2008}.
\newblock \bibinfo{title}{Bayesian image reconstruction based on voronoi
  diagrams}.
\newblock \bibinfo{journal}{The Astrophysical Journal} \bibinfo{volume}{672},
  \bibinfo{pages}{1272--1285}.
%Type = Article
\bibitem[{Candan et~al.(2000)Candan, Kutay and Ozaktas}]{fft}
\bibinfo{author}{Candan, C.}, \bibinfo{author}{Kutay, M.},
  \bibinfo{author}{Ozaktas, H.}, \bibinfo{year}{2000}.
\newblock \bibinfo{title}{The discrete fractional fourier transform}.
\newblock \bibinfo{journal}{Signal Processing, IEEE Transactions on}
  \bibinfo{volume}{48}, \bibinfo{pages}{1329--1337}.
%Type = Article
\bibitem[{{Casassus} et~al.(2006){Casassus}, {Cabrera}, {F{\"o}rster},
  {Pearson}, {Readhead} and {Dickinson}}]{firstVersionMEM}
\bibinfo{author}{{Casassus}, S.}, \bibinfo{author}{{Cabrera}, G.F.},
  \bibinfo{author}{{F{\"o}rster}, F.}, \bibinfo{author}{{Pearson}, T.J.},
  \bibinfo{author}{{Readhead}, A.C.S.}, \bibinfo{author}{{Dickinson}, C.},
  \bibinfo{year}{2006}.
\newblock \bibinfo{title}{{Morphological Analysis of the Centimeter-Wave
  Continuum in the Dark Cloud LDN 1622}}.
\newblock \bibinfo{journal}{The Astrophysical Journal} \bibinfo{volume}{639},
  \bibinfo{pages}{951--964}.
%Type = Article
\bibitem[{Casassus et~al.(2015)Casassus, Marino, Perez, Roman, Dunhill,
  Armitage, Cuadra, Wootten, van~der Plas, Cieza, Moral, Christiaens and
  Montesinos}]{co65}
\bibinfo{author}{Casassus, S.}, \bibinfo{author}{Marino, S.},
  \bibinfo{author}{Perez, S.}, \bibinfo{author}{Roman, P.},
  \bibinfo{author}{Dunhill, A.}, \bibinfo{author}{Armitage, P.J.},
  \bibinfo{author}{Cuadra, J.}, \bibinfo{author}{Wootten, A.},
  \bibinfo{author}{van~der Plas, G.}, \bibinfo{author}{Cieza, L.},
  \bibinfo{author}{Moral, V.}, \bibinfo{author}{Christiaens, V.},
  \bibinfo{author}{Montesinos, M.}, \bibinfo{year}{2015}.
\newblock \bibinfo{title}{Accretion kinematics through the warped transition
  disk in hd142527 from resolved co(6–5) observations}.
\newblock \bibinfo{journal}{The Astrophysical Journal} \bibinfo{volume}{811},
  \bibinfo{pages}{92}.
%Type = Article
\bibitem[{Chen(2011)}]{illposed}
\bibinfo{author}{Chen, W.}, \bibinfo{year}{2011}.
\newblock \bibinfo{title}{The ill-posedness of the sampling problem and
  regularized sampling algorithm}.
\newblock \bibinfo{journal}{Digit. Signal Process.} \bibinfo{volume}{21},
  \bibinfo{pages}{375--390}.
%Type = Book
\bibitem[{Cheng et~al.(2014)Cheng, Grossman and McKercher}]{cudapro}
\bibinfo{author}{Cheng, J.}, \bibinfo{author}{Grossman, M.},
  \bibinfo{author}{McKercher, T.}, \bibinfo{year}{2014}.
\newblock \bibinfo{title}{Professional CUDA C Programming}.
\newblock EBL-Schweitzer, \bibinfo{publisher}{Wiley}.
%Type = Inproceedings
\bibitem[{Clark(1999)}]{clark}
\bibinfo{author}{Clark, B.G.}, \bibinfo{year}{1999}.
\newblock \bibinfo{title}{{Coherence in Radio Astronomy}}, in:
  \bibinfo{editor}{Taylor, G.B.}, \bibinfo{editor}{Carilli, C.L.},
  \bibinfo{editor}{Perley, R.A.} (Eds.), \bibinfo{booktitle}{ASP Conf. Ser.
  180: Synthesis Imaging in Radio Astronomy II}, pp.~\bibinfo{pages}{1+}.
%Type = Article
\bibitem[{{Cornwell}(1988)}]{cornwell}
\bibinfo{author}{{Cornwell}, T.J.}, \bibinfo{year}{1988}.
\newblock \bibinfo{title}{{Radio-interferometric imaging of very large
  objects}}.
\newblock \bibinfo{journal}{Astronomy and Astrophysics} \bibinfo{volume}{202},
  \bibinfo{pages}{316--321}.
%Type = Article
\bibitem[{{Cornwell} and {Evans}(1985a)}]{smemda}
\bibinfo{author}{{Cornwell}, T.J.}, \bibinfo{author}{{Evans}, K.F.},
  \bibinfo{year}{1985}a.
\newblock \bibinfo{title}{{A simple maximum entropy deconvolution algorithm}}.
\newblock \bibinfo{journal}{AAP} \bibinfo{volume}{143},
  \bibinfo{pages}{77--83}.
%Type = Article
\bibitem[{{Cornwell} and {Evans}(1985b)}]{mem1}
\bibinfo{author}{{Cornwell}, T.J.}, \bibinfo{author}{{Evans}, K.F.},
  \bibinfo{year}{1985}b.
\newblock \bibinfo{title}{{A simple maximum entropy deconvolution algorithm}}.
\newblock \bibinfo{journal}{Astronomy and Astrophysics} \bibinfo{volume}{143},
  \bibinfo{pages}{77--83}.
%Type = Inproceedings
\bibitem[{{Coughlan} and {Gabuzda}(2013)}]{astromem2}
\bibinfo{author}{{Coughlan}, C.P.}, \bibinfo{author}{{Gabuzda}, D.C.},
  \bibinfo{year}{2013}.
\newblock \bibinfo{title}{{Imaging VLBI polarimetry data from Active Galactic
  Nuclei using the Maximum Entropy Method}}, in: \bibinfo{booktitle}{European
  Physical Journal Web of Conferences}, p. \bibinfo{pages}{07009}.
%Type = Article
\bibitem[{Donoho et~al.(1992)Donoho, Jhonstone, Koch and Stern}]{nearlyBlack}
\bibinfo{author}{Donoho, D.L.}, \bibinfo{author}{Jhonstone, I.M.},
  \bibinfo{author}{Koch, J.C.}, \bibinfo{author}{Stern, A.S.},
  \bibinfo{year}{1992}.
\newblock \bibinfo{title}{Maximum entropy and the nearly black object}.
\newblock \bibinfo{journal}{Journal of the Royal Statistical Society}
  \bibinfo{volume}{54}.
%Type = Inproceedings
\bibitem[{Gregg and Hazelwood(2011)}]{Gregg2011}
\bibinfo{author}{Gregg, C.}, \bibinfo{author}{Hazelwood, K.},
  \bibinfo{year}{2011}.
\newblock \bibinfo{title}{Where is the data? why you cannot debate {CPU} vs.
  {GPU} performance without the answer}, in: \bibinfo{booktitle}{(IEEE-ISPASS)
  IEEE International Symposium on Performance Analysis of System and Software},
  \bibinfo{publisher}{{IEEE}}.
%Type = Article
\bibitem[{Gull and Daniell(1978)}]{NatureMEM}
\bibinfo{author}{Gull, S.F.}, \bibinfo{author}{Daniell, G.J.},
  \bibinfo{year}{1978}.
\newblock \bibinfo{title}{Image reconstruction from incomplete and noisy data}.
\newblock \bibinfo{journal}{Nature} \bibinfo{volume}{272},
  \bibinfo{pages}{686--690}.
%Type = Article
\bibitem[{Hogbom(1974)}]{hogbom}
\bibinfo{author}{Hogbom, J.A.}, \bibinfo{year}{1974}.
\newblock \bibinfo{title}{{Aperture Synthesis with a Non-Regular Distribution
  of Interferometer Baselines}}.
\newblock \bibinfo{journal}{Astron. Astrophys. Suppl. Ser.}
  \bibinfo{volume}{15}, \bibinfo{pages}{417--426}.
%Type = Article
\bibitem[{{Lannes} et~al.(1997){Lannes}, {Anterrieu} and {Marechal}}]{persuit}
\bibinfo{author}{{Lannes}, A.}, \bibinfo{author}{{Anterrieu}, E.},
  \bibinfo{author}{{Marechal}, P.}, \bibinfo{year}{1997}.
\newblock \bibinfo{title}{{Clean and Wipe}}.
\newblock \bibinfo{journal}{Astronomy and Astrophysicss} \bibinfo{volume}{123},
  \bibinfo{pages}{183--198}.
%Type = Article
\bibitem[{{Lochner} et~al.(2015){Lochner}, {Natarajan}, {Zwart}, {Smirnov},
  {Bassett}, {Oozeer} and {Kunz}}]{BIRO}
\bibinfo{author}{{Lochner}, M.}, \bibinfo{author}{{Natarajan}, I.},
  \bibinfo{author}{{Zwart}, J.T.L.}, \bibinfo{author}{{Smirnov}, O.},
  \bibinfo{author}{{Bassett}, B.A.}, \bibinfo{author}{{Oozeer}, N.},
  \bibinfo{author}{{Kunz}, M.}, \bibinfo{year}{2015}.
\newblock \bibinfo{title}{{Bayesian inference for radio observations}}.
\newblock \bibinfo{journal}{Monthly Notices of the Royal Astronomical Society}
  \bibinfo{volume}{450}, \bibinfo{pages}{1308--1319}.
%Type = Article
\bibitem[{Marechal and Wallach(2009)}]{marechal}
\bibinfo{author}{Marechal, P.}, \bibinfo{author}{Wallach, D.},
  \bibinfo{year}{2009}.
\newblock \bibinfo{title}{Fourier synthesis via partially finite convex
  programming}.
\newblock \bibinfo{journal}{Mathematical and Computer Modelling}
  \bibinfo{volume}{49}, \bibinfo{pages}{2206 -- 2212}.
%Type = Inproceedings
\bibitem[{McMullin et~al.(2007)McMullin, Waters, Schiebel, Young and
  Golap}]{casa}
\bibinfo{author}{McMullin, J.P.}, \bibinfo{author}{Waters, B.},
  \bibinfo{author}{Schiebel, D.}, \bibinfo{author}{Young, W.},
  \bibinfo{author}{Golap, K.}, \bibinfo{year}{2007}.
\newblock \bibinfo{title}{Astronomical data analysis software and systems xvi},
  in: \bibinfo{editor}{Shaw, R.A.}, \bibinfo{editor}{Hill, F.},
  \bibinfo{editor}{Bell, D.J.} (Eds.), \bibinfo{booktitle}{ASP Conf. Ser.},
  \bibinfo{address}{San Francisco, CA}. p. \bibinfo{pages}{127}.
%Type = Article
\bibitem[{{Narayan} and {Nityananda}(1986)}]{MEM}
\bibinfo{author}{{Narayan}, R.}, \bibinfo{author}{{Nityananda}, R.},
  \bibinfo{year}{1986}.
\newblock \bibinfo{title}{{Maximum entropy image restoration in astronomy}}.
\newblock \bibinfo{journal}{Annual Review of Astronomy and Astrophysics}
  \bibinfo{volume}{24}, \bibinfo{pages}{127--170}.
%Type = Article
\bibitem[{Neff et~al.(2015)Neff, Eilek and Owen}]{astromem1}
\bibinfo{author}{Neff, S.G.}, \bibinfo{author}{Eilek, J.A.},
  \bibinfo{author}{Owen, F.N.}, \bibinfo{year}{2015}.
\newblock \bibinfo{title}{The complex north transition region of centaurus a:
  Radio structure}.
\newblock \bibinfo{journal}{The Astrophysical Journal} \bibinfo{volume}{802},
  \bibinfo{pages}{87}.
%Type = Book
\bibitem[{Nesterov(2004)}]{ConvexOptimization}
\bibinfo{author}{Nesterov, Y.}, \bibinfo{year}{2004}.
\newblock \bibinfo{title}{Introductory Lectures on Convex Optimization}.
\newblock \bibinfo{publisher}{Springer-Verlag}, \bibinfo{address}{New York, NY,
  USA}.
%Type = Book
\bibitem[{Nocedal and Wright(2006)}]{numericaloptimization}
\bibinfo{author}{Nocedal, J.}, \bibinfo{author}{Wright, S.J.},
  \bibinfo{year}{2006}.
\newblock \bibinfo{title}{Numerical Optimization}.
\newblock \bibinfo{publisher}{Springer-Verlag}, \bibinfo{address}{New York, NY,
  USA}.
%Type = Book
\bibitem[{{NVIDIA Corporation}(2016a)}]{cudabest}
\bibinfo{author}{{NVIDIA Corporation}}, \bibinfo{year}{2016}a.
\newblock \bibinfo{title}{CUDA C Best Practices Guide}.
\newblock \bibinfo{publisher}{NVIDIA Corporation}.
\newblock \bibinfo{note}{Version 8.0}.
%Type = Book
\bibitem[{{NVIDIA Corporation}(2016b)}]{cudaprofiler}
\bibinfo{author}{{NVIDIA Corporation}}, \bibinfo{year}{2016}b.
\newblock \bibinfo{title}{CUDA Profiler User's Guide}.
\newblock \bibinfo{publisher}{NVIDIA Corporation}.
\newblock \bibinfo{note}{Version 8.0}.
%Type = Misc
\bibitem[{{OpenMP Architecture Review Board}(2015)}]{openmp}
\bibinfo{author}{{OpenMP Architecture Review Board}}, \bibinfo{year}{2015}.
\newblock \bibinfo{title}{{OpenMP} openmp application programming interface
  version 4.5}.
%Type = Article
\bibitem[{Perkins et~al.(2015)Perkins, Marais, Zwart, Natarajan and
  Smirnov}]{montblanc}
\bibinfo{author}{Perkins, S.}, \bibinfo{author}{Marais, P.},
  \bibinfo{author}{Zwart, J.}, \bibinfo{author}{Natarajan, I.},
  \bibinfo{author}{Smirnov, O.}, \bibinfo{year}{2015}.
\newblock \bibinfo{title}{Montblanc: Gpu accelerated radio interferometer
  measurement equations in support of bayesian inference for radio
  observations}.
\newblock \bibinfo{journal}{Astronomy and Computing} \bibinfo{volume}{12},
  \bibinfo{pages}{73--85}.
%Type = Article
\bibitem[{{Pina} and {Puetter}(1993)}]{1993Pixon}
\bibinfo{author}{{Pina}, R.K.}, \bibinfo{author}{{Puetter}, R.C.},
  \bibinfo{year}{1993}.
\newblock \bibinfo{title}{{Bayesian image reconstruction - The pixon and
  optimal image modeling}}.
\newblock \bibinfo{journal}{Publication of the Astronomical Society of the
  Pacific} \bibinfo{volume}{105}, \bibinfo{pages}{630--637}.
%Type = Article
\bibitem[{Pinte et~al.(2016)Pinte, Dent, Menard, Hales, Hill, Cortes and
  de~Gregorio-Monsalvo}]{hltau1}
\bibinfo{author}{Pinte, C.}, \bibinfo{author}{Dent, W.R.F.},
  \bibinfo{author}{Menard, F.}, \bibinfo{author}{Hales, A.},
  \bibinfo{author}{Hill, T.}, \bibinfo{author}{Cortes, P.},
  \bibinfo{author}{de~Gregorio-Monsalvo, I.}, \bibinfo{year}{2016}.
\newblock \bibinfo{title}{Dust and gas in the disk of hl tauri: Surface
  density, dust settling, and dust-to-gas ratio}.
\newblock \bibinfo{journal}{The Astrophysical Journal} \bibinfo{volume}{816},
  \bibinfo{pages}{25}.
%Type = Book
\bibitem[{Press et~al.(1992)Press, Teukolsky, Vetterling and
  Flannery}]{numericalrecipes}
\bibinfo{author}{Press, W.H.}, \bibinfo{author}{Teukolsky, S.A.},
  \bibinfo{author}{Vetterling, W.T.}, \bibinfo{author}{Flannery, B.P.},
  \bibinfo{year}{1992}.
\newblock \bibinfo{title}{Numerical Recipes in C (2Nd Ed.): The Art of
  Scientific Computing}.
\newblock \bibinfo{publisher}{Cambridge University Press},
  \bibinfo{address}{New York, NY, USA}.
%Type = Article
\bibitem[{Quinn et~al.(2015)Quinn, Axelrod, Bird, Dodson, Szalay and
  Wicenec}]{ska}
\bibinfo{author}{Quinn, P.}, \bibinfo{author}{Axelrod, T.},
  \bibinfo{author}{Bird, I.}, \bibinfo{author}{Dodson, R.},
  \bibinfo{author}{Szalay, A.}, \bibinfo{author}{Wicenec, A.},
  \bibinfo{year}{2015}.
\newblock \bibinfo{title}{Delivering {SKA} science}.
\newblock \bibinfo{journal}{CoRR} \bibinfo{volume}{abs/1501.05367}.
%Type = Article
\bibitem[{{Rastorgueva, E. A.} et~al.(2011){Rastorgueva, E. A.}, {Wiik, K. J.},
  {Bajkova, A. T.}, {Valtaoja, E.}, {Takalo, L. O.}, {Vetukhnovskaya, Y. N.}
  and {Mahmud, M.}}]{memGeneralized}
\bibinfo{author}{{Rastorgueva, E. A.}}, \bibinfo{author}{{Wiik, K. J.}},
  \bibinfo{author}{{Bajkova, A. T.}}, \bibinfo{author}{{Valtaoja, E.}},
  \bibinfo{author}{{Takalo, L. O.}}, \bibinfo{author}{{Vetukhnovskaya, Y. N.}},
  \bibinfo{author}{{Mahmud, M.}}, \bibinfo{year}{2011}.
\newblock \bibinfo{title}{Multi-frequency vlba study of the blazar s5 0716+714
  during the active state in 2004.}
\newblock \bibinfo{journal}{Astronomy and Astrophysics} \bibinfo{volume}{529}.
%Type = Article
\bibitem[{{Rau} and {Cornwell}(2011)}]{MULTIFREQUENCY}
\bibinfo{author}{{Rau}, U.}, \bibinfo{author}{{Cornwell}, T.J.},
  \bibinfo{year}{2011}.
\newblock \bibinfo{title}{{A multi-scale multi-frequency deconvolution
  algorithm for synthesis imaging in radio interferometry}}.
\newblock \bibinfo{journal}{Astronomy and Astrophysics} \bibinfo{volume}{532},
  \bibinfo{pages}{A71}.
%Type = Article
\bibitem[{Sutton and Wandelt(2006)}]{optimal_image}
\bibinfo{author}{Sutton, E.C.}, \bibinfo{author}{Wandelt, B.D.},
  \bibinfo{year}{2006}.
\newblock \bibinfo{title}{Optimal image reconstruction in radio
  interferometry}.
\newblock \bibinfo{journal}{The Astrophysical Journal Supplement Series}
  \bibinfo{volume}{162}, \bibinfo{pages}{401}.
%Type = Article
\bibitem[{Tamayo et~al.(2015)Tamayo, Triaud, Menou and Rein}]{hltau2}
\bibinfo{author}{Tamayo, D.}, \bibinfo{author}{Triaud, A.H.M.J.},
  \bibinfo{author}{Menou, K.}, \bibinfo{author}{Rein, H.},
  \bibinfo{year}{2015}.
\newblock \bibinfo{title}{Dynamical stability of imaged planetary systems in
  formation: Application to hl tau}.
\newblock \bibinfo{journal}{The Astrophysical Journal} \bibinfo{volume}{805},
  \bibinfo{pages}{100}.
%Type = Article
\bibitem[{{Tamura} et~al.(2015){Tamura}, {Oguri}, {Iono}, {Hatsukade},
  {Matsuda} and {Hayashi}}]{sdp}
\bibinfo{author}{{Tamura}, Y.}, \bibinfo{author}{{Oguri}, M.},
  \bibinfo{author}{{Iono}, D.}, \bibinfo{author}{{Hatsukade}, B.},
  \bibinfo{author}{{Matsuda}, Y.}, \bibinfo{author}{{Hayashi}, M.},
  \bibinfo{year}{2015}.
\newblock \bibinfo{title}{{High-resolution ALMA observations of SDP.81. I. The
  innermost mass profile of the lensing elliptical galaxy probed by 30
  milli-arcsecond images}}.
\newblock \bibinfo{journal}{Publications of the Astronomical Society of Japan}
  \bibinfo{volume}{67}, \bibinfo{pages}{72}.
%Type = Book
\bibitem[{Taylor et~al.(1999)Taylor, Carilli and Perley}]{libroAstro2}
\bibinfo{editor}{Taylor, G.B.}, \bibinfo{editor}{Carilli, C.L.},
  \bibinfo{editor}{Perley, R.A.} (Eds.), \bibinfo{year}{1999}.
\newblock \bibinfo{title}{{Synthesis Imaging in Radio Astronomy II}}. volume
  \bibinfo{volume}{180} of \textit{\bibinfo{series}{Astronomical Society of the
  Pacific Conference Series}}.
\newblock \bibinfo{publisher}{Astronomical Society of the Pacific},
  \bibinfo{address}{San Francisco}.
%Type = Article
\bibitem[{Temlyakov(2008)}]{greedyAlgorithm}
\bibinfo{author}{Temlyakov, V.N.}, \bibinfo{year}{2008}.
\newblock \bibinfo{title}{{Greedy approximation}}.
\newblock \bibinfo{journal}{Acta Numerica} \bibinfo{volume}{17},
  \bibinfo{pages}{235--409}.
%Type = Book
\bibitem[{Thompson et~al.(2008)Thompson, Moran and Swenson}]{libroAstro}
\bibinfo{author}{Thompson, A.}, \bibinfo{author}{Moran, J.},
  \bibinfo{author}{Swenson, G.}, \bibinfo{year}{2008}.
\newblock \bibinfo{title}{Interferometry and Synthesis in Radio Astronomy}.
\newblock \bibinfo{publisher}{Wiley}.
%Type = Article
\bibitem[{{van Haarlem} et~al.(2013){van Haarlem}, {Wise}, {Gunst}
  et~al.}]{lofar}
\bibinfo{author}{{van Haarlem}, M.P.}, \bibinfo{author}{{Wise}, M.W.},
  \bibinfo{author}{{Gunst}, A.W.}, et~al., \bibinfo{year}{2013}.
\newblock \bibinfo{title}{{LOFAR: The LOw-Frequency ARray}}.
\newblock \bibinfo{journal}{Astronomy \& Astrophysics} \bibinfo{volume}{556},
  \bibinfo{pages}{A2}.
%Type = Article
\bibitem[{{Warmuth} and {Mann}(2013)}]{astromem3}
\bibinfo{author}{{Warmuth}, A.}, \bibinfo{author}{{Mann}, G.},
  \bibinfo{year}{2013}.
\newblock \bibinfo{title}{{Thermal and nonthermal hard X-ray source sizes in
  solar flares obtained from RHESSI observations. I. Observations and
  evaluation of methods}}.
\newblock \bibinfo{journal}{Astronomy and Astrophysics} \bibinfo{volume}{552},
  \bibinfo{pages}{A86}.
%Type = Article
\bibitem[{{Winkel} et~al.(2016){Winkel}, {Lenz} and {Fl{\"o}er}}]{gridding}
\bibinfo{author}{{Winkel}, B.}, \bibinfo{author}{{Lenz}, D.},
  \bibinfo{author}{{Fl{\"o}er}, L.}, \bibinfo{year}{2016}.
\newblock \bibinfo{title}{{Cygrid: A fast Cython-powered convolution-based
  gridding module for Python}}.
\newblock \bibinfo{journal}{Astronomy and Astrophysics} \bibinfo{volume}{591},
  \bibinfo{pages}{A12}.

\end{thebibliography}
